\newif\ifsubmode
\submodefalse

\newif\ifprintfig
\printfigtrue

\newif\ifemulate
\emulatetrue

\ifsubmode
  \documentclass[12pt,preprint]{aastex}
  \received{}
  \accepted{}
  \journalid{}{}
  \articleid{}{}
\else
   \documentclass{emulateapj}
   \submitted{{\it Submitted to AJ}}
\fi

\newcommand{\kms}{\,km~s$^{-1}$}

\def\spose#1{\hbox to 0pt{#1\hss}}
\def\simlt{\mathrel{\spose{\lower 3pt\hbox{$\mathchar"218$}}
     \raise 2.0pt\hbox{$\mathchar"13C$}}}
\def\simgt{\mathrel{\spose{\lower 3pt\hbox{$\mathchar"218$}}
     \raise 2.0pt\hbox{$\mathchar"13E$}}}

\shorttitle{``Galaxy,'' defined}
\shortauthors{Willman \& Strader}

\begin{document}

\title{``Galaxy,'' defined}

\author{B.\ Willman\altaffilmark{1}, J.Strader\altaffilmark{2}}
\altaffiltext{1}{Haverford College, Department of Astronomy, 370 Lancaster Avenue, Haverford, PA, 19041, bwillman@haverford.edu}
\altaffiltext{2}{Harvard-Smithsonian CfA, 60 Garden St., Cambridge, MA
02144, jstrader@cfa.harvard.edu}

\begin{abstract}
\renewcommand{\thefootnote}{\fnsymbol{footnote}} 

A growing number of low luminosity and low surface brightness
astronomical objects challenge traditional notions of both galaxies
and star clusters.  To address this challenge, we propose a definition
of galaxy that does not depend on a cold dark matter model of the
universe: A galaxy is a gravitationally bound collection of stars
whose properties cannot be explained by a combination of baryons and
Newton's laws of gravity. After exploring several possible
observational diagnostics of this definition, we critically examine
the classification of ultra-faint dwarfs, globular clusters,
ultra-compact dwarfs, and tidal dwarfs. While kinematic studies
provide an effective diagnostic of the definition in many regimes,
they can be less useful for compact or very faint systems. To explore
the utility of using the [Fe/H] spread as a complementary diagnostic,
we use published spectroscopic [Fe/H] measurements of 16 Milky Way
dwarfs and 24 globular clusters to uniformly calculate their [Fe/H]
spreads and associated uncertainties.  Our principal results are: (i)
no known, old star cluster less luminous then $M_{V} = -10$ has a
significant ($\gtrsim$0.1 dex) spread in its iron abundance; (ii)
known ultra-faint dwarf galaxies can be unambiguously classified with
a combination of kinematic and [Fe/H] observations; (iii) the observed
[Fe/H] spreads in massive ($\ga 10^{6} M_{\odot}$) globular clusters
do not necessarily imply that they are the stripped nuclei of dwarfs,
nor a need for dark matter; and (iv) if ultra-compact dwarf galaxies
reside in dark matter halos akin to those of ultra-faint dwarfs of the
same half-light radii, then they will show no clear dynamical
signature of dark matter.  We suggest several measurements that may
assist the future classification of massive globular clusters,
ultra-compact dwarfs, and ultra-faint galaxies.  Our galaxy definition
is designed to be independent of the details of current observations
and models, while our proposed diagnostics can be refined or replaced
as our understanding of the universe evolves.


\end{abstract}


\section{Introduction}\label{intro_sec}
\renewcommand{\thefootnote}{\fnsymbol{footnote}}

It has been nearly a century since the term ``galaxy''
\citep{desitter17,crommelin18a,shapley19a} was first applied to the
spiral nebulae that were later found to be true extra-galactic stellar
systems (\citealt{slipher17a,hubble26a}, see also \citealt{graham11a}
and references therein).  The term ``cluster'' has been used to refer
to Milky Way (MW) open and globular cluster star systems since their
initial discoveries more than 200 years ago \citep{messier1781}.  In
the early 20th century, the primary differences between objects
classified as galaxies and star clusters were (i) the
milky\footnote{The word ``galaxy'' derives directly from the Greek
  word for ``milky''.} appearances of galaxies versus the grainy
appearances of star clusters, and (ii) the ``island universe''
environments of galaxies versus the association of star clusters with
the MW system.  In the intervening years, galaxies and star clusters
have largely been classified based on their physical sizes: galaxies
have typical sizes of hundreds of pc to tens of kpc whereas star
clusters have typical sizes of a few pc, with a scatter to tens of pc.

The lion's share of known star clusters and galaxies can be classified
by this simple ``I know it when I see it'' size-based distinction.
However, there are a growing number of astronomical objects that are
not so easily classified---those at extreme low luminosities and
surface brightnesses, and those filling gaps previously observed in
size--luminosity space \citep{misgeld11a}.  These objects currently
hold the most promise to shed new light on galaxy formation at the
bottom of the hierarchy and the distribution of dark matter to the
smallest possible size and mass scales.  For example, ultra-compact
dwarfs (UCDs) have luminosities ($-13 < M_{V} < -9$) similar to those
of both dwarf spheroidal galaxies and luminous globular clusters
(GCs), but sizes (10 pc $<$ $r_{half}$ $<$ 100 pc) intermediate to
both populations and dynamical mass-to-light ratios of $\sim
2-5$\footnote{All mass-to-light ratios are in $V$ unless otherwise
  stated.}, larger than those typical of GCs
\citep[e.g.,][]{hilker99a,drinkwater03a,hasegan05a,evstigneeva07a,mieske08a,chilingarian11a,brodie11a}.
Some ultra-faint MW satellites (e.g., Segue 1, Segue 2, Bo\"otes II,
Willman 1) have also challenged our notion of galaxies.  These objects
have luminosities ($-3 < M_{V} < -1$) lower than those of nearly any
known old star cluster or dwarf galaxy, physical sizes (20 pc $<$
$r_{half}$ $<$ 40 pc) between those of most star clusters and dwarf
galaxies, and dynamical mass-to-light ratios as high as 3000
\citep{simon11a}.  While tidal dwarfs\footnote{We use ``tidal dwarf''
  rather than ``tidal dwarf galaxy'' throughout this paper, to avoid
  presupposing a galaxy definition for this class of objects.} may
provide a piece to this puzzle, they have proved difficult to study
and classify themselves (see \citealt{duc12a} for a review).

The origin and properties of systems such as UCDs, extreme MW dwarf
satellites, and tidal dwarfs are fundamental to many open questions in
galaxy formation and cosmology.  They might hold unique clues to
relationships between different classes of hot stellar systems (e.g.,
giant elliptical galaxies, dwarf elliptical galaxies, dwarf spheroidal
galaxies, UCDs, nuclear star clusters, GCs;
\citealt{dabringhausen08a,wolf10a,misgeld11a,zaritsky11a}).  They
might be our best luminous tracers of sub-galactic dark matter.
Having a well-defined classification scheme will be essential to these
studies, since imminent and upcoming wide-field surveys, including
Pan-STARRS 1 \citep{kaiser02a}, the Southern Sky Survey
\citep{keller07a}, the Dark Energy Survey \citep{DES} and LSST
\citep{ivezic08a}, are expected to reveal large numbers of previously
unseen low surface-brightness systems.

As the rate of discoveries and the diversity of the known cosmic zoo
increases, the question ``What is a galaxy?''  is being discussed at
conferences and in the literature
\citep[e.g.,][]{gilmore07a,kroupa08a,vandenbergh08a,forbes11a}.  The
most common distinction currently made between galaxies and star
clusters is the presence of dark matter---galaxies reside at the
centers of dark matter halos and star clusters do not
\citep[e.g.,][]{simon11a,tollerud11a,willman11a}.  One strength of this
definition is that it facilitates studies of dwarf galaxies in a
cosmological context: it allows a straightforward connection between
the set of objects classified as galaxies and a comparison with the
predictions of dark matter plus galaxy formation models.  One weakness
of this physically motivated definition is the fact that the cold dark
matter model is a theory rather than a physical law.

Other recently proposed definitions for a galaxy have included
$r_{half}$ $>$ 100 pc, a relaxation time longer than a Hubble time, or
complex stellar populations \citep{forbes11a}.  Although each of these
definitions have their own strengths (namely that they are
straightforward to diagnose), they each also have
shortcomings\footnote{See also the related, informal discussion linked
  at the URL \url{http://arxiv.org/abs/1101.3309.}}.  For example,
size-based classifications are becoming increasingly arbitrary as
size--luminosity space is becoming continuously populated with objects
\citep{misgeld11a}.  The concept of ``complex'' stellar populations
has also become ill-defined now that light element abundance spreads
have been identified in a large fraction of MW GCs
\citep[e.g.,][]{gratton04a,cohen05a,dantona05a,carretta09b}, which
were once thought to be pristine examples of simple stellar
populations.

In this paper, we tweak past definitions of the term ``galaxy'' with
the aim of appealing to a sufficiently broad cross-section of
astronomers that consensus might be reached.  In \S2, we motivate the
importance of having a clear definition of galaxy within the
astronomical community, and then present a physically motivated
definition.  In \S 3, we consider kinematics and complementary
indirect diagnostics such as [Fe/H] spread and population-based
diagnostics that might be used to test whether an object is a galaxy.
In \S4 we examine the known properties of ultra-faint dwarfs, UCDs,
GCs, and tidal dwarfs in the context of our proposed definition.

\section{Galaxy Definition}

The classification of astronomical objects is more than a semantic
pursuit.  Words matter.  The terminology we choose to describe our
research affects how appealing and accessible it sounds to funding
agencies and the public.  Words also dramatically affect the research
community's ability to draw global conclusions from diverse sets of
astronomical objects.  For example, massive star cluster $\omega$Cen
may be the remnant of a stripped, nucleated dwarf galaxy
\citep{lee99a,bekki03a}.  However, it is cataloged as a GC and thus
is not considered in studies of the MW's dwarf galaxy population.
Well-defined, well-chosen classification schemes therefore improve our
understanding of the universe by facilitating meaningful comparisons
between models and observations.  Conversely, ill-defined, ill-chosen
classification schemes can muddy our understanding of astrophysical
phenomena.

Although no single definition of galaxy will be optimal for all
purposes, we propose a physically motivated definition that will
facilitate studies of galaxies both in and out of a cosmological
context:

\begin{center}

  {\it A galaxy is a gravitationally bound collection of stars whose
    properties cannot be explained by a combination of baryons and
    Newton's laws of gravity.}

\end{center}

In a dark matter context (whether cold, warm, self-interacting, or
other), this definition loosely translates to measuring whether an
object contains dark matter.  Alternatively, the definition can be
interpreted to delineate those objects for which non-standard theories
of gravity could be relevant \citep[e.g.,][]{milgrom83, sotiriou10}.
For such theories, our definition would require distinct observable
consequences of non-standard gravity to be imprinted on an
astrophysical system for it to be classified as a galaxy.  In the
interest of simplicity, we focus Sections 3 and 4 of this paper on
diagnosing our proposed galaxy definition in a dark matter-based
context.  However, we encourage others to more explicitly explore
these diagnostics in alternative contexts.

We refer to Newton's laws on a macroscopic scale; this part of the
definition should be considered to include objects that require
general relativity to be understood. We recognize that galaxies can
have a significant amount of baryonic mass in non-stellar forms such
as gas and dust. It is uncertain whether the local universe harbors
any ``dark galaxies'' that contain gas but are entirely free of stars
\citep{minchin05,duc08a}.  Whether these objects should be classified as
galaxies is an open question that we do not confront here but that
deserves further debate.

A purely descriptive astronomical classification (such as relaxation
time) may be relatively straightforward for either observers or
theorists to implement.  A weakness of the definition proposed here
thus lies in finding adequate diagnostics to measure whether the
properties of the lowest luminosity and most compact objects are
explicable with baryons and Newton's Laws (see \S 3 for a discussion
of possible diagnostics).

Because galaxy formation itself is hierarchical in a cold dark matter
universe, there is no trivial distinction between a single galaxy with
satellite galaxies (such as the MW) and a galaxy group or cluster. As
a diagnostic, \citet{busha12a} propose that a bound collection of
galaxies is a ``galaxy'' rather than a galaxy cluster if at least 50\%
of the stellar light is associated with one central object. This
diagnostic is ultimately driven by the decreasing efficiency of galaxy
formation in more massive dark matter halos.

The \citet{busha12a} diagnostic helps guide intuition when exercising
reasonable common sense in applying our definition to astrophysical
systems.  For example, the intracluster star population of a galaxy
cluster is composed of a gravitationally bound collection of stars
whose dynamics cannot be explained by orbits within a Newtonian
potential well dominated by cluster baryons.  Such a system should not
be classified as a galaxy because it is physically associated with a
galaxy cluster.  Similarly, the Milky Way's stellar halo is a merely a
component of the Milky Way---not its own galaxy.

\section{Galaxy Diagnostics}

In this and in the following section, we focus on objects for which a
galaxy classification (or lack thereof) tends to be ambiguous.

\subsection{Stellar kinematics}

The most direct way to determine whether an object contains dark
matter, or whether its properties are otherwise inconsistent with
Newtonian gravity, is to conduct a kinematic study.  The present day
mass of a system is typically derived from its kinematics using
formalism based on Newton's laws of gravity and the assumption of
dynamical equilibrium.  This dynamical mass can then be compared with
the total mass present in the form of stars, stellar remnants, and
gas.  If dynamical mass exceeds the baryonic mass, then dark matter
must be present or one of the dynamical assumptions---such as
Newtonian gravity or virial equilibrium---must be flawed.

There are many regimes in which dynamical studies can be translated
with few assumptions into Newtonian masses
\citep[e.g.,][]{walker09c,wolf10a}.  \citet{wolf10a} showed that the
half-light mass of a dispersion supported system could be robustly
calculated with only mild assumptions about the orbital anisotropy
of its constituent stars.  They derive $M_{half}$ =
4G$^{-1}$$<\sigma^2_{los}>$$r_{half}$.  Here $M_{half}$ is the
total mass within the 3D deprojected half-light radius,
$<\sigma^2_{los}>$ is the luminosity weighted square of the line of
sight velocity dispersion, and $r_{half}$ is the 2D projected
half-light radius.  Such calculations have yielded $(M/L)_{half}$
as high as $\sim$3000 for a MW satellite galaxy \citep[Segue
1,][]{simon11a}.  

It is not always possible to diagnose a galaxy definition based on
dynamical $(M/L)_{half}$ alone.  Many authors have looked at the
relationship between $M/L$ and other system properties (such as
luminosity, see e.g. Figures 3 and 5 in \citealt{tollerud11a} and
Figure 4 in \citealt{wolf10a}.)  While typical star clusters stand out
as having low $(M/L)_{half}$ ($\sim$ 1-5) for their luminosities ($L
\sim$ 10$^{4-6} L_{\odot}$), dispersion supported galaxies ($L \sim$
10$^{8-10} L_{\odot}$) have similar $(M/L)_{half}$ as star clusters.
In such cases, a combination of $(M/L)_{half}$ and other population
arguments may be used to diagnose a galaxy classification (see also
\S3.3).  Alternatively, dynamical modeling including tracers at larger
distances can reveal $M/L$ outside of $r_{half}$.

If the existence of dark matter is the correct interpretation of
galaxy dynamics, then dynamical classification of galaxies may be
robust to the effect of tidal mass loss.  Simulations show that
galaxies tidally stripped of mass should maintain a high dynamical
mass-to-light ratio.  For example, \citet{penarrubia08a} showed that
the mass-to-light ratios of tidally evolving dwarf galaxies increase
over time, assuming they reside in cuspy dark matter halos.  Even if
the dark matter halos hosting dwarf galaxies are cored, their central
dark matter density slopes remain constant during tidal evolution
\citep{penarrubia10a}. 

\subsubsection{Special Considerations}

Generally, dynamical $M/L$ $\gtrsim$ 5 may be taken to diagnose a
galaxy classification, because such $M/L$ are larger than expected
from typical stellar populations.  However, a number of challenges
face attempts to determine whether an observed dynamical $M/L$ of a
system is consistent with expectations from baryons alone - especially
for systems with $M/L \lesssim$ 10, low intrinsic velocity dispersions, or
low surface brightness. Some of these challenges are fairly obvious, as 
the dynamical $M/L$ expected for a purely
baryonic population varies significantly with: age, metallicity,
initial mass function, dynamical state, and gas content.  In this
section, we highlight several specific examples which are less
commonly discussed in the literature.  See also \S4.2.1 for a more
nuanced discussion of dynamical $M/L$ in the context of UCDs.

Several effects can cause an overestimate in dynamical mass, and thus
an overestimate of $M/L$.  For example, the orbital motions of binary
stars can inflate a system's observed velocity dispersion.  A recent,
multi-epoch velocity study of Segue 1 suggests that binaries should
not pose a major problem for the dynamical classification of systems
with intrinsic velocity dispersions of at least a few \kms
\citep{martinez11a,simon11a}.  However, binaries do materially impact
lower velocity dispersion systems \citep{bradford11a}, and models
based on more extreme assumptions than previously considered identify
regions of parameter space where binaries could impact Segue 1-like
velocity dispersion systems \citep{mcconnachie10a}.  Tidally unbound
and MW foreground stars can also contaminate spectroscopic samples of
a MW companion's stars and inflate its observed velocity dispersion.
The effect of such contaminants can be mitigated by a combination of
careful simulations of the MW foreground and its
color--magnitude--velocity distribution \citep{willman11a}, the use of
spectroscopic abundance indicators, statistical approaches to
identifying object members \citep[e.g.,][]{walker09b}, and approaches
to eliminating tidally stripped stars that have been informed by
N-body simulations \citep{klimentowski07}.

Other effects may alternatively cause an underestimate of stellar
mass, and thus an overestimate of the presence of non-stellar mass.
For example, \citet{hernandez12a} shows that the lowest luminosity
systems ($L$ $\sim$ 500 $L_{\odot}$) can have $(M/L)_{stellar}$
between 1 and 10 simply from the stochastic effects of sampling an IMF
with a small number of stars.  A tidally stripped, dynamically relaxed
(and therefore mass-segregated) GC can also have a super-stellar $M/L$
once the majority of its mass has been lost.  Models of star cluster
evolution that include the effects of mass segregation and the
Galaxy's tidal field have shown that high fractions of stellar
remnants accumulate in the center as a cluster is stripped of mass
\citep{vesperini97,giersz01a,baumgardt03a}.  Although possible, it
should be rare to observe a system so tidally stripped that its global
$M/L$ is significantly inflated by this mechanism.  For example,
although Palomar 5 is estimated to be $\sim$100 Myr from complete
destruction (less than 1\% of its total lifetime), it is observed to
have $M/L_{dyn}$ $<$ 1 \citep{odenkirchen02a,dehnen04}.  Observational
limitations may also generate ambiguity in the dynamical
classification of the lowest luminosity ($L$ $<$ 1000 $L_{\odot}$) and
low velocity dispersion ($<$3 \kms) systems.  For example, Segue 3 (L
$= 90^{+90}_{-40}$ $L_{\odot}$, $d$ $\sim$ 17 kpc) contains only a few
dozen member stars brighter than $r = 22$.  32 of Segue 3's stars were
observed with Keck/DEIMOS to obtain velocity measurements with
uncertainties per exposure per star of $\sim$3$-$10 \kms~
\citep{fadely11a}.  With a $\sigma_{los}$ of 0.3\kms~expected based on
stars and Newtownian gravity alone, its measured velocity dispersion
of 1.2 $\pm$ 2.6 \kms~is dynamically consistent with either a galaxy
or a star cluster interpretation.  Even with techniques which retrieve
stellar velocities from medium-resolution spectra with uncertainties
$<$1 \kms \citep{koposov11a}, star-poor systems need to reside within
$\sim$20 kpc for there to be a sufficient number of stars bright
enough to spectroscopically observe with high S/N with a 10m-class
telescope.

\subsection{[Fe/H] Spread}

Another way to directly constrain the potential well in which a system
formed is the presence of an [Fe/H] spread.  The use of [Fe/H] as a
diagnostic for our proposed galaxy definition is supported by a
combination of models of supernova winds in low-mass systems and the
observed abundances of stars in well-studied dwarfs and GCs.  Iron is
produced by supernovae (both Type II and Ia), so a dispersion in
[Fe/H] implies that the system was able to retain supernova ejecta to
form multiple generations of stars. The energetic winds of supernovae
can only be retained in a gravitational well of sufficient depth.
Estimates for the GC mass needed to retain SN ejecta are
$>$few$\times$10$^6$ M$_{\odot}$
\citep[e.g.,][]{dopita86,baumgardt08b}.  Observed [Fe/H] spreads of
over 1 dex combined with inferred stellar masses of $\sim$1000
M$_{\odot}$ or less have thus contributed to a galaxy classification
for both Segue 1 and Willman 1
\citep{martin07a,norris10a,simon11a,willman11a}.

\subsubsection{Calculating $\sigma_{\rm [Fe/H]}$}

To empirically investigate the difference in [Fe/H] spread,
$\sigma_{\rm [Fe/H]}$, between well-studied dwarf galaxies and GCs, we
estimate the spread and associated uncertainty for each of 16 dwarfs
and 24 GCs with publicly available, spectroscopic [Fe/H] measurements.
We only used [Fe/H] measurements based on actual iron lines, rather
than studies that infer iron abundance from the calcium triplet or
photometry.  We used Bayesian Markov Chain Monte Carlo techniques to
fit a normal distribution to the stellar [Fe/H] values for each
object, taking into account the reported measurement uncertainties and
assuming flat priors.\footnote{The relevant (simple) code is available
  on request.}  We summarize the standard deviation of each sample,
$\sigma_{\rm [Fe/H]}$, as the median of its posterior distribution,
together with a 68.2\% credible interval (analogous to the usual
$1\sigma$ confidence interval).  Calculated values of [Fe/H],
$\sigma_{\rm [Fe/H]}$, associated uncertainties, and references are
summarized in Table 1.  The uncertainties on the variances are an
increasing function of decreasing sample size, because small samples
poorly sample the underlying [Fe/H] distribution.

A few notes on unusual cases: For Segue 1, we included the star from
\citet{simon11a} with only an upper limit to its [Fe/H] as a censored
datum in our analysis. We used the largest set of [Fe/H] values for
$\omega$Centauri \citep{johnson10a}. However, this sample is biased
against the most metal-rich subpopulation because it is
magnitude-limited in $V$.  We thus consider our estimate of its [Fe/H]
spread to be a lower limit. The \citet{marino11a} data for M22 does
not contain uncertainties, and so our reported $\sigma_{\rm [Fe/H]}$
is an upper limit.  Our analysis does not include the Terzan 5 GC
despite claims of an [Fe/H] spread in this object
\citep{ferraro09a,origlia11a}, owing to its $\sim$ solar abundance
(and thus different origin than the old metal-poor stellar populations
we are primarily considering) and the possibility that the sample may
be partially contaminated by bulge stars.  We also did not include NGC
5824, in which \citet{saviane12a} have reported $\sigma_{\rm [Fe/H]}
\sim 0.11 - 0.14$ dex, because this measurement is based on the
Calcium triplet (thus revealing a Ca spread, not necessarily an Fe
spread).  The GC  NGC~2419 is known to display a $\sim$0.2 dex spread
in Ca, but none in Fe \citep{cohen10a}.

Although reasonable indicators of the dispersion in [Fe/H], the values
in Table 1 should be considered with caution before comparing in
detail with models.  The accuracy of our estimates of the variance of
[Fe/H] (and its uncertainty) rely on (i) the appropriateness of the
underlying Gaussian model, (ii) clean membership samples, and (iii)
accurate uncertainties for individual stars.  For the fainter dwarfs
in this set the first condition rarely holds
\citep[e.g.,][]{kirby11a}, so our estimated variances should be taken
as indicators of the spread in metallicity rather than as exact
values.  The faintest dwarfs may also have a small number of
interloper stars in their membership samples (see also \S3.2.4).  The
third condition---estimating accurate uncertainties---is most relevant
for GCs, because their measured $\sigma_{\rm [Fe/H]}$ are comparable
to (or less than) the measurement uncertainties for single stars. For
this paper, we have included the random uncertainty in the \ion{Fe}{1}
abundance as the standard error of the mean, while \citet{carretta09a}
included no measurement uncertainties in their calculation of [Fe/H]
spread. The practical effect is that the intrinsic [Fe/H] spreads we
derive for GCs in this paper are slightly smaller than those in
\citet{carretta09a}, by typically 0.01 dex.  Like \citet{carretta09a},
we emphasize that their and our values are upper limits to be true
[Fe/H] spreads because of our limited ability to model the full
measurement uncertainties on each star (see \citealt{carretta09a}
for detailed discussion of the relevant modeling issues).

\subsubsection{$\sigma_{\rm [Fe/H]}$ in M$_{\rm V} \gtrsim -10$ Objects}

\begin{figure*}[t!]
\epsscale{0.9}
\plotone{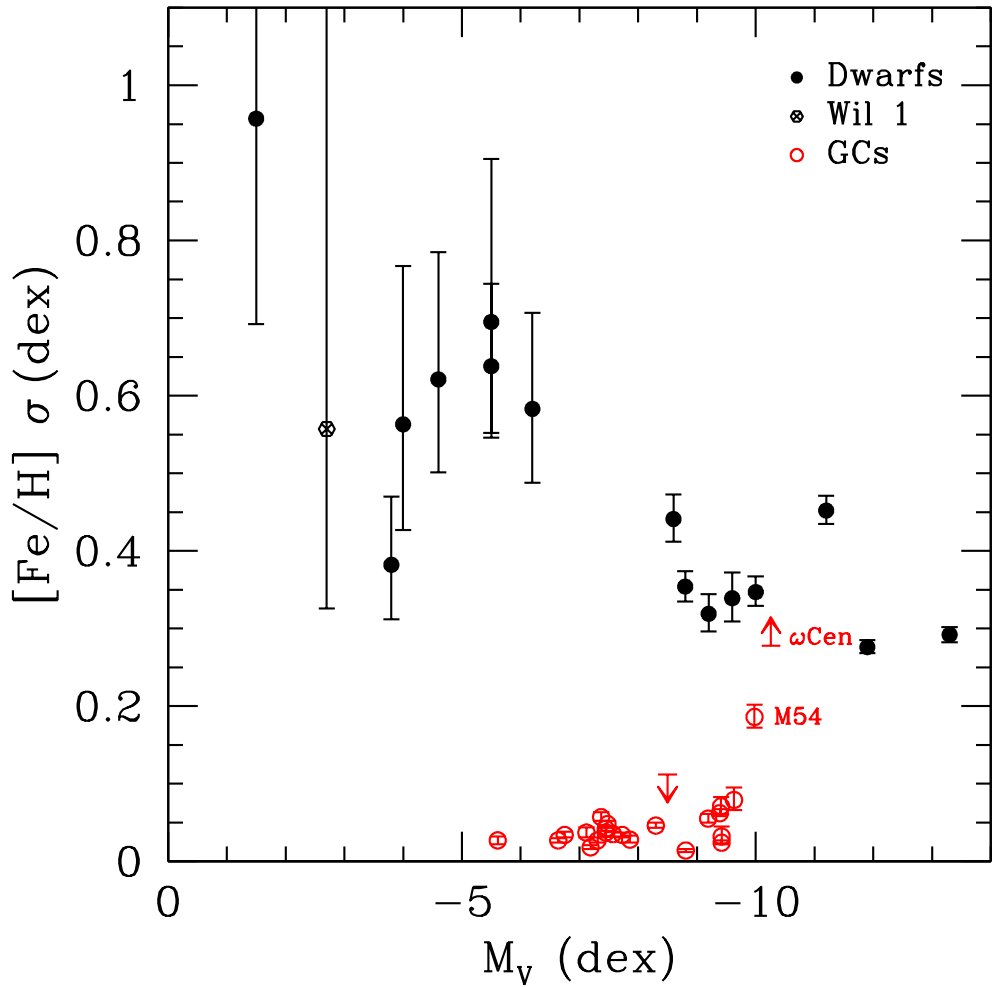}
 \caption{The dispersion in [Fe/H] measured for MW dwarf galaxies
   (black, filled) and
   globular clusters (red, open), calculated assuming an underlying Gaussian
   distribution.  The systems shown with dwarf galaxy symbols in this
   figure have been independently classified as
   galaxies by dynamical studies.  Willman 1 does not have a definitive
   dynamical classification, and so is shown as an open hexagon with a
   cross.  By this figure, a galaxy classification can be indirectly inferred
   from Willman 1's spread in [Fe/H].  The presence of a spread in
   [Fe/H] can diagnose a galaxy definition because it constrains
   the depth of the potential well in which a system formed, as supernova ejecta must be retained to form further
   generations of stars.  Error bars show the estimated uncertainty on each
   dispersion given the [Fe/H] measurement uncertainties on the
   individual member stars. Values and references are summarized in
   Table 1.  Figure 7 of \citet{carretta10b} shows a
   figure similar in spirit to this, but for a smaller set of objects and
   without measurement uncertainties.
   \label{fig_FeH}}
\end{figure*}

Figure~\ref{fig_FeH} shows $\sigma_{\rm [Fe/H]}$ for MW dwarf galaxies
(filled, black circles) and MW GCs (open red circles) as a function of
absolute magnitude.  Uncertainty bars show the 68.2\% confidence
intervals.  We show objects with dynamical classifications of galaxy
or star cluster as different symbols in Figure~\ref{fig_FeH} to
highlight the regime in which $\sigma_{\rm [Fe/H]}$ results in the
same inference about a system's potential well as a dynamical study.
This tests whether $\sigma_{\rm [Fe/H]}$ may be used as a galaxy
diagnostic in cases where dynamical studies are inconclusive.

This figure shows a striking difference between the [Fe/H] spreads
observed for dwarf galaxies and GCs.  The dwarfs all have spreads of
0.3--0.7 dex (even higher for Segue 1), whereas none of the GCs less
luminous than M$_{V} = -10$ have substantial [Fe/H] dispersions.
After the upper limit of $\sigma_{\rm [Fe/H]}=$ 0.1 dex estimated for
M22, the next highest spread is 0.08 dex estimated for NGC
6441. Although these values are small, they are formally greater than
zero with $> 99$\% probability (as calculated above). These estimates
may reflect the detection of minor star-to-star variations in [Fe/H]
in GCs less luminous than $M_{V} = -10$. However, in light of the
caveats given above, they may yet be found to be consistent with no
star-to-star variation in [Fe/H].

For objects less luminous than $M_{V} = -10$, the dichotomy between
$\sigma_{\rm [Fe/H]}$ of dwarf galaxies and GCs underscores that the
dwarf galaxies formed within much deeper potential wells than the GCs.
We conclude that a $\sigma_{\rm [Fe/H]} > 0.2$ dex in such systems
would be sufficient to diagnose a galaxy classification because it
would not be explicable with a combination of baryons and Newtonian
gravity (without invoking substantial mass loss).  While iron is not
the only element that may provide relevant insight to the
gravitational potential wells of objects in this luminosity regime, it
is clear that iron spread provides a powerful diagnostic of the
provenance of such objects.

\subsubsection{$\sigma_{\rm [Fe/H]}$ in M$_{V} \lesssim -10$ Objects}

The interpretation of the $\sigma_{\rm [Fe/H]}$ spreads observed in
the two GCs more luminous than $M_{V} = -10$, M54 and $\omega$Cen, is
less straightforward.  One interpretation of the spreads in M54 and
$\omega$Cen is that they are the nuclear star cluster cores remaining
from a stripped dwarf galaxy (M54: Sgr core, \citealt{sarajedini95a};
$\omega$Cen, \citealt{lee99a,bekki03a}). It remains to be seen whether
the properties of the gravitationally bound remains of such a stripped
galaxy would satisfy our definition of a galaxy, and be formally
classified as such.  Recent observations have discovered a significant
amount of tidal debris that may be associated with $\omega$Cen
\citep{maj12}. Sgr is already a classified galaxy, so M54 would not be
considered a separate entity.

An alternative interpretation of the [Fe/H] spreads in these $M_{star}
> 10^6$ $M_{\odot}$ clusters is self-enrichment by SNe without the
additional gravitational influence of dark matter or a non-Newtonian
effect.  This interpretation is complicated by the fact that M54 and
$\omega$Cen do not actually have the highest escape velocities of the
GCs in our sample.  Using the fitted relation between central velocity
dispersion and central escape velocity, $v_{esc,0}/\sigma_0 = 3.7 +
0.9(c-1.4)$, from \citet{gnedin02a}, we find that 8 of the 62 GCs with
velocity dispersions reported in the 2010 edition of
\citet{harrisGCcat} have central escape velocities larger than M54's
$v_{esc,0} \sim 45$ \kms (not including $\omega$Cen).  5 of these (47
Tuc, NGC 2808, NGC 6388, NGC 6441, and M15) are in our sample and do
not display [Fe/H] spreads $\ga 0.1$ dex.  NGC 6441 and 6388 have
escape velocities of 72 and 76 \kms, respectively, larger than
$\omega$Cen's escape velocity of 61 \kms. A caveat to this analysis is
that these values are measured at the present day.  At earlier times,
these GCs were all more massive but have since undergone stellar
evaporation and tidal mass loss; some may have also had different
sizes. All of these factors could have affected their relative
abilities to retain supernova ejecta.

Observations of GCs in other galaxies provide tentative support for
self-enrichment in iron in $M_V < -$10 GCs. For example, HST/ACS
photometry of three of the most massive GCs in M31 are suggestive of
spreads in [Fe/H] on the red giant branch
\citep{fuentescarrera08a}. The dynamical masses of these GCs range
from $2-6 \times 10^{6} M_{\odot}$, comparable to or larger than
$\omega$ Cen \citep{strader11a}. The M31 cluster G1 ($3 \times 10^{6}
M_{\odot}$) also may have a significant [Fe/H] spread
\citep{meylan01a}.

Separately, a number of groups have identified evidence of self-enrichment
in extragalactic GCs. Precise photometry of blue, metal-poor GCs in a
variety of galaxies
\citep{harris06a,mieske06a,strader06a,spitler06a,forbes10b,mieske10a} shows a
correlation between magnitude and color for metal-poor GCs. This
mass--metallicity relationship is not observed in all galaxies
studied, but a typical relation is $Z \sim M^{0.4}$, where $Z$ is the
mean metallicity of the GC and $M$ is its mass. The onset of the
correlation appears to be between $\sim 2\times 10^{5}$ and $10^{6} 
M_{\odot}$. The slope and onset mass of the correlation can be
reasonably explained by models in which the GCs self-enrich in iron
\citep{bailin09a,strader08a}.

If (nearly) all GCs with stellar masses above few $\times 10^6$
M$_{\odot}$ display [Fe/H] spreads, then it is likely these spreads
accrue from self-enrichment without the help of an additional
gravitational field.  More extensive spectroscopic and photometric
campaigns to quantify the [Fe/H] spreads of extragalactic GCs will be
essential to develop a fuller picture of the connection between
$\sigma_{\rm [Fe/H]}$ and the formation channel(s) of objects with
$M_{\rm V} < -10$.

\subsubsection{A Relationship Between $\sigma_{\rm [Fe/H]}$ and
  M$_{V}$ For Dwarfs?}

Figure~\ref{fig_FeH} displays another striking trend in addition to
the dwarf/GC dichotomy: the apparent increase in $\sigma_{\rm [Fe/H]}$
with decreasing luminosity (see also \S6.2 of \citealt{kirby11a}).
While the dispersion in [Fe/H] for most MW dwarf galaxies with $M_{V}
< -8$ (the classical dwarfs) is $0.3-0.4$ dex, the dispersion for most
of the lower luminosity dwarfs (the ultra-faint dwarfs) is $0.5-0.6$
dex.  The most likely explanations for this apparent trend are: (i) a
true physical difference in the $\sigma_{\rm [Fe/H]}$ of the least
luminous systems, (ii) a systematic bias in the calculated
$\sigma_{\rm [Fe/H]}$ as the model assumptions become increasingly
poor with decreasing luminosity, or (iii) a result of a low level of
foreground contamination that disproportionately affects spectroscopic
samples of the lowest surface brightness systems.  The faintest dwarfs
have tails at the metal-richer ends of their metallicity distribution
functions that are not present in the classical dwarfs.  It is not yet
clear whether those metal-richer tails are physical or a result of
mild contamination in the samples.  Exploring the relative likelihood
of these three scenarios is beyond the scope of this paper, but will
be imperative to pursue in the future.

\subsection{Indirect Diagnostics: Population Arguments}

Population arguments rely on the assumption of a single classification
for all astrophysical objects known to populate a particular region of
parameter space.  Such arguments are handy because, for example, it
would be both impractical and unnecessary to conduct a detailed
analysis of each of the 200 million galaxies cataloged by the eighth
Sloan Digital Sky Survey data release \citep{dr8} before classifying
them as such.  The most common population-based classification is
simply the size-based classification that is naturally made for
galaxies with scale size $\gtrsim$ 1 kpc.  All objects satisfying this
``I know it when I see it'' criterion that have been studied in
sufficient detail have been kinematically shown to satisfy our
proposed definition of galaxy (not including tidal dwarfs, see \S4.4.)
Some kinematic studies of galaxies have postulated that no unseen
matter or modification of Newtonian gravity may be needed to explain
their dynamics \citep[e.g.,][]{romanowsky03a}.  However, such studies
have always been shown to be flawed on theoretical grounds
\citep[e.g.,][]{dekel05a} or were refuted by subsequent observational
studies.

Attempts have been made to connect, or distinguish, galaxies and star
clusters using scaling relations that combine their metallicities,
effective sizes, internal velocities, luminosities or derivatives
thereof.  Such studies have recently focused on variants of the
Fundamental Plane such as the Fundamental Manifold
\citep{forbes11b,zaritsky11a} and the Fundamental Curve
\citep{tollerud11a}.  These scalings reveal similarities and
differences in the ways baryons coalesce within different types of
systems.  However, the scalings do not seem to shed light on the
classification of objects as a star cluster or a dwarf galaxy, in a
way more meaningful than $M/L$ within $r_{half}$
\citep[e.g.,][]{forbes08a,tollerud11a,zaritsky11a}.  One simple
difference between galaxies and globular clusters as a population is
the metallicity-luminosity relation observed for galaxies (but not
star clusters, UCDs, or nearby tidal dwarfs) over a wide range of
stellar masses \citep[e.g.][]{skillman89a,tremonti04,woo08a,kirby11a}.
Although the metallicity and luminosity of an individual object would
not be sufficient to classify it as a galaxy or star cluster,
consideration of the metallicities and luminosities of a population of
objects may aid in its classification (see also \S4.2.2).  It is also
worthwhile to consider the placement of individual ambiguous objects
with respect to observed scaling relations.  Inconsistency with
well-established relationships on a case-by-case basis may be a sign
that some of the cautions raised in \S3.1.1 are affecting the
kinematics, effective mass, or size measured for an object.

Another approach to population-based classification is to include a
broad set of properties such as spatial distribution, metallicity, and
orbits when looking for subtle trends within a diverse set of
observables.  The combination of such a set of clues may help reveal
whether some object or type of object with an ambiguous classification has an
origin (and thus, possibly, classification) more similar to that of
star clusters or of dwarf galaxies.  \citet{brodie11a} recently
conducted a thorough analysis of UCDs around M87 in the Virgo cluster.
They combined size--luminosity, age--metallicity, spatial
distribution, and orbital dynamics to infer the possible co-existence
in size and luminosity of three sub-populations of UCDs: the stripped
nuclei of dEs, remnants from more massive red galaxies (either their
nuclei or merged clusters), and genuine star clusters.
 
Although we do not aim to be exhaustive, throughout \S4 we will
mention some specific indirect diagnostics that may contribute to a
galaxy classification.

\section{Some Examples Throughout the Cosmic Zoo}

In this section, we use the diagnostics in \S3 to consider the
classification of four populations of astrophysical objects: extreme
ultra-faint dwarf galaxies, UCDs, GCs, and tidal dwarfs.

\subsection{Ultra-faint dwarfs with r$_{ half} <$ 50 pc}

We begin our discussion with extreme ultra-faint dwarfs, because their
classification is starting to converge in the literature.  The term
``ultra-faint dwarfs'' refers to the dwarf galaxies with absolute
magnitudes fainter than $M_{V} \sim -8$. Currently, such objects are
only known around the Milky Way and M31 because they are difficult to
detect, although the Next Generation Virgo Cluster Survey should soon
reveal them in Virgo.  The most extreme of these objects (Segue 1,
Segue 2, Bo\"otes II, Willman 1) are observed to have $M_{V} \sim
-2.5$ and $r_{half} \sim$ 30 pc.

These extreme objects have total luminosities less than individual
bright red giant branch stars.  Their sizes are intermediate between
typical GCs and low luminosity dwarf spheroidal galaxies.  Despite
their extreme and unusual properties, direct and/or indirect
diagnostics support a galaxy definition for all four of these objects.
With an $(M/L)_{half}$ $\sim$ 3400 and $\sigma_{\rm [Fe/H]}=$
0.75$^{+0.42}_{-0.23}$ dex (Table~1), Segue 1 is a galaxy as diagnosed
by both its kinematic and its $\sigma_{\rm [Fe/H]}$.  Taken at face
value, the dynamics of Willman 1's stars require a high dynamical mass
relative to its stellar mass.  However, its irregular kinematic
distribution hinders drawing a robust classification from kinematics
alone \citep{willman11a}.  Regardless, the substantial spread in
[Fe/H] among three member stars in Willman 1 (0.56$^{+0.58}_{-0.23}$
dex, Table~1) demonstrates its galaxy classification.  A dynamical
study based on small numbers of stars in Segue 2 \citep{belokurov09a}
is consistent with a galaxy classification, although the uncertainties
are still large.  Finally, tidal arguments for Bo\"otes II have
suggested that it may need a substantial dark matter component for it
to be self-bound \citep{walsh08a}.

It is essential that all (candidate) extreme ultra-faint dwarfs close
enough to study with 10m-class telescopes are spectroscopically
investigated.  Surveys like DES and LSST have the potential to uncover
large numbers of objects like Segue 1 to distances beyond the reach of
today's spectroscopic resources.  If a sufficient number of nearby
Segue 1-like objects are demonstrated to be galaxies, then systems
discovered to share that region of size--luminosity space in the
future might be classified as galaxies without extensive
follow-up. Even now, it is not yet certain whether Segue 2 and
Bo\"otes II should be counted as galaxies or remnants thereof.
Their classifications will greatly impact the predicted number of
luminous dwarfs orbiting the Milky Way and our (currently minimal)
knowledge of the bottom of the galaxy luminosity function.  

As a population, the MW's ultra-faint dwarfs follow
luminosity-metallicity and luminosity-($M/L$) relations
\citep[e.g.,][]{geha09a}.  These scaling relations rule out
pathological explanations for the ultra-faints as a population, such
as clumps in tidal streams or stellar streams at orbital apocenter.
When in doubt for any particular object, hypothesis testing against
the spatial-kinematic predictions of a specific model can be used to
effectively vet a galaxy classification.  For example,
\citet{zolotov11a} showed that the highly elliptical Hercules dwarf
spheroidal is inconsistent with a cusp catastrophe hypothesis.

\subsection{UCDs} 

Like ``ultra-faint dwarf'', the term UCD has no formal definition.  It
is generally used to refer to systems with $-13 \lesssim M_{V}
\lesssim -9$ and 10 pc $<$ $r_{half}$ $<$ 100 pc.  This population of
objects has proved particularly challenging to classify.  With up to
100 UCDs possibly orbiting M87 alone \citep{brodie11a}, whether or not
these should be counted as galaxies bears great importance for
understanding the dwarf galaxy population of the Virgo Cluster in a
cosmological context.  Thus far, studies seem to be converging on the
conclusion that multiple formation channels may be required to explain
the UCDs as a population, such as very massive star clusters or as the
stripped nuclei of dwarf galaxies
\citep{brodie11a,chiboucas11a,chilingarian11a,darocha11a}.  In this
section, we do not review the work relying on population arguments or
detailed studies of individual UCDs
\citep[e.g.,][]{maraston04a,fellhauer05a,norris11b} to reach this
conclusion.  We instead discuss the efficacy of UCD kinematic studies
in a cosmological context and consider possible future kinematic and
[Fe/H] UCD classification diagnostics.

\subsubsection{UCD Kinematics}

Dynamical studies of UCDs do not provide a clear diagnosis of a galaxy
classification.  UCD3, the most luminous UCD in the Fornax cluster
(M$_{V}$ = -13.55, $r_{half}$ = 87 pc), is the only UCD with spatially
resolved kinematics \citep{frank11a}.  UCD3 has less than a 33\% mass
contribution from dark matter within 200 pc, and $M/L$ = 3.6$\pm$0.3,
if it is assumed that mass follows light.  This $M/L$ may be consistent
with the $M/L$ of its stellar population (\citealt{chilingarian11a},
however see \citealt{mieske06a} and \citealt{firth09a} who estimate a
lower stellar $M/L$).  The spatially unresolved dynamical studies of
other UCDs yield dynamical $M/L$ = $2-5$, plausibly (but not
certainly) consistent with their stellar $M/L$
\citep[e.g.,][]{hilker99a,drinkwater03a,hasegan05a,evstigneeva07a,mieske08a,chilingarian11a}.
The dynamical $M/L$ of Virgo cluster UCDs seem to be systematically
higher than Fornax cluster UCDs \citep{mieske08a}.  The inflated Virgo
UCD $M/L$s may be explained by unusual IMFs; a top-heavy IMF can yield
a large fraction of dark stellar remnants
\citep{dabringhausen09a,dabringhausen12a}, while a bottom-heavy IMF is
rich in low-mass M dwarfs with high individual $M/L$.

It is not surprising that dynamical studies of UCDs do not easily
yield a galaxy classification, even if (for example) they do presently
reside in dark matter halos.  To quantify this, we must begin with a
reasonable hypothesis for the amount of dark matter expected within
the half-light radii of UCDs if they reside in dark matter halos.
There are no simulations of sufficient spatial resolution to predict
the expected amount of dark matter in the innermost $\sim$30 pc of a
dark matter halo, and the highest resolution simulations do not
include the effect of baryons, star formation, or feedback.  Moreover,
there are known differences between the central mass densities
observed for dwarf galaxies and the central dark matter densities
predicted for dwarf galaxies using dark matter only simulations
\citep{boylankolchin11a,boylankolchin12a} that have, in some cases,
been resolved with baryonic physics
\citep[e.g.,][]{governato10a,pontzen12a}.  Similarly,
\citet{tollerud11a} find that the observed central mass densities of
UCDs are not consistent with residing in the Navarro-Frenk-White
profile dark matter halos predicted by dark matter simulations.

We therefore rely on an empirical hypothesis for the possible
dark matter content of UCDs: they contain the same amount of dark
matter within their half-light radii as known dwarf galaxies with the
same half-light radii.  We consider two MW dwarfs: Segue 1
\citep[$\sigma_{los}$ = 3.7$^{+1.4}_{-1.1}$ \kms, $M_{V}$ =
-1.5$^{+0.6}_{-0.8}$, $r_{half}$= 29 $\pm$ 6 pc]{martin08b,simon11a}
and Coma Berenices \citep[M$_{V}$ = -3.6$\pm$0.6, $r_{half}$= 74
$\pm$ 4 pc]{simon07a,munoz10a}.  Using the \citet{wolf10a} formula, we
calculate $M_{Segue1,half}$ = 3.7$^{+2.9}_{-2.3} \times$ 10$^5$
$M_{\odot}$ and $M_{ComBer,half}$ = 1.5$\pm$0.5$\times$ 10$^6$
$M_{\odot}$.  To obtain the half-light dark matter masses of these
objects, we simply subtract out their approximate stellar masses
assuming a stellar $M/L$ of 2.  Because Segue 1 and Coma Berenices are
highly dark matter dominated, the derived dark matter masses depend
little on our assumed value of $(M/L)_{star}$.

\begin{figure*}[t!]
\epsscale{0.8}
\plotone{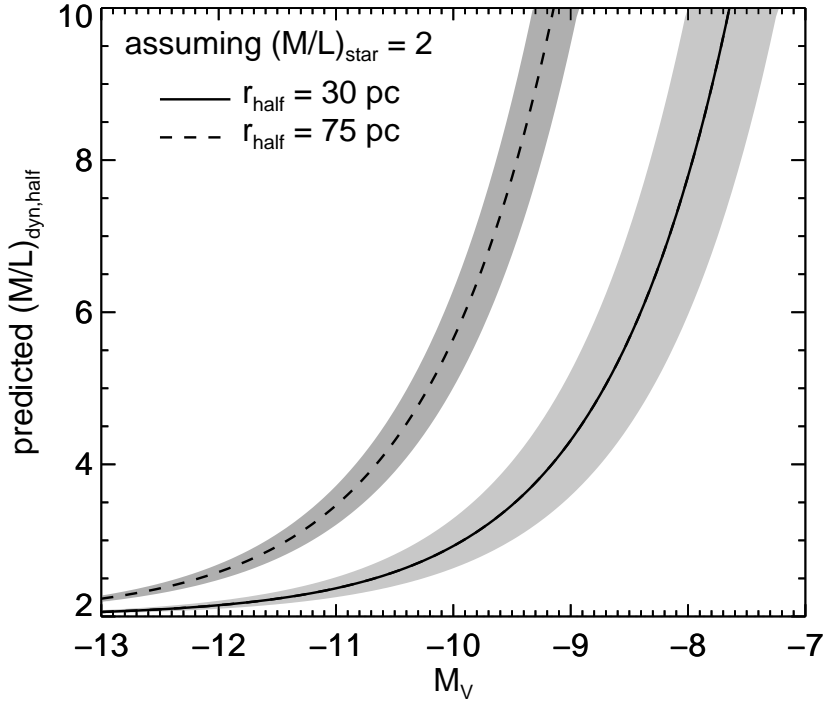}
\caption{The predicted $(M/L)_{half}$ of UCDs with $r_{half}$ =
  30 or 75 pc, assuming they reside in dark matter halos like those
  inferred for Segue 1 and ComBer, respectively.  Typical UCDs should
  not display dynamical evidence for dark matter, even if they do
  reside in the centers of dark matter halos.
  \label{fig_MtoL}}
\end{figure*}

We use the half-light dark matter masses of Segue 1 and Coma Berenices
to predict the possible dynamical $(M/L)_{half}$ of UCD-luminosity
systems with half-light radii of 30 pc and 75 pc.
Figure~\ref{fig_MtoL} shows the resulting predictions, as a function
of absolute magnitude and assuming a stellar $M/L$ = 2 for the UCDs.
UCDs are typically observed to have $-13 <$ $M_{V} < -9$ and 10 pc $<$
$r_{half}$ $<$ 100 pc \citep[see
e.g.][]{madrid10a,brodie11a,misgeld11b}.  We predict that UCDs in dark
halos would have dynamical $M/L$ within their half-light radii of
$2-3$, consistent with observations.  Given the large uncertainties in
deriving stellar $M/L$, this prediction confirms that dynamics will
not be able to unambiguously reveal the presence of dark matter in
most individual UCDs.  Less luminous UCDs have less baryonic mass, and
so will be more dynamically affected by the presence of dark matter if
they reside in halos similar to those of more luminous UCDs.  We also
predict that among UCDs of similar luminosity, those with larger
scale-sizes should have systematically higher dark matter fractions.
This prediction makes sense, because larger half-light radii enclose a
larger fraction of an object's dark matter halo, if UCDs of similar luminosity
reside in similar dark matter halos.  Current observations do not
bear a clear signature of this predicted relationship
\citep{mieske08a}.  However, because of possible system-to-system variations and
uncertainties in stellar $M/L$, it is impossible (to date) to draw robust
conclusions about the dynamical evidence for dark matter or lack
thereof.
 
In making the quantitative predictions in Figure~2, we have assumed
that UCDs contain the same amount of dark matter within their
half-light radii as known dwarf galaxies with the same half-light
radii.  The dark matter halos inhabited by UCDs may instead have
higher mass density than those inhabited by MW ultra-faint dwarfs,
owing to gravitational contraction.  Alternatively they may have lower
mass density, owing to the far greater amount of feedback from star
formation and death experienced by UCDs with orders of magnitude more
stars than ultra-faint dwarfs.  Nevertheless, Figure~2 demonstrates a
reasonable model in which dark matter is not dynamically detectable in
most UCDs, but may be detectable in the least luminous UCDs.  The
relationship we predict between half-light radius and dynamical mass
is dependent only on the assumption that similar luminosity UCDs
inhabit similar dark matter halos.

\subsubsection{UCD [Fe/H]}

Even if it is possible to assess $\sigma_{\rm [Fe/H]}$ in UCDs with
$M_{V} < -10$, it would not easily aid in their classification (see
\S3.2).  \citet{brodie11a} have recently argued that objects with
lower stellar masses are also part of the UCD population around M87 in
Virgo.  NGC 2419, a MW GC, has a size (21 pc, \citealt{harrisGCcat})
and absolute magnitude ($M_{V}$ = -9.42, \citealt{harrisGCcat})
consistent with the lower luminosity UCDs around M87.  At face value,
NGC 2419's lack of an [Fe/H] spread (Table~1) suggests that star
clusters may form with the sizes and luminosities of at least some
UCDs.  However, NGC 2419's spread in Ca ($\sim$0.2 dex) may be
difficult to reconcile with the inferred depth of its potential well.
Unlike spreads in lighter elements, a Ca spread might require
enrichment by supernovae \citep{cohen10a}.

It would be extremely interesting if future studies could measure (or
set limits on) the [Fe/H] spread of a set of lower luminosity UCDs to
see whether they all lack a spread in [Fe/H], as observed for typical
star clusters.  Another hint to a possible UCD--dwarf galaxy
connection---or lack thereof---may be their average [Fe/H].  UCDs fall
above the metallicity-luminosity relationship followed by dwarf
galaxies \citep[e.g.,][see also discussion in \S3.3]{chilingarian11a}.
If the UCDs are stripped remnants of nucleated dwarfs then they once
would have been more luminous and may have fallen on observed
metallicity-luminosity relationships.

\subsection{Globular Clusters}  

A combination of dynamics, $\sigma_{\rm [Fe/H]}$, and several indirect
diagnostics show that GCs, as a population, do not satisfy our
definition of galaxy and do not presently inhabit dark matter halos.
We briefly discuss this evidence here, because we should neither take
for granted that canonical GCs do not satisfy our proposed definition
of galaxy, nor take for granted that they should be ignored in efforts
to map dark matter substructure around the MW and other galaxies.  For
example, the spatial distribution of MW halo GCs is consistent with
the predicted present-day distribution of early forming dark matter
peaks \citep{brodie06a,moore06a}.  This similarity could be
interpreted as evidence that GCs themselves reside in the center of
present day dark matter halos and, if so, should be included in
studies that rely on dwarf galaxies as luminous tracers of the spatial
and mass distribution of dark matter.

No dynamical study of GCs has yielded a dynamical mass in excess of
stellar mass, even for lower surface density \citep[Palomar
13,][]{bradford11a} and tidally disrupting clusters \citep[Palomar
5,][]{odenkirchen02a}.  In light of the dynamical arguments presented
for UCDs, GCs would be unlikely to exhibit straightforward dynamical
evidence for dark matter even if they did reside in dark matter halos.
The [Fe/H] analysis in \S3.2 and shown in Figure~1 instead provides
direct evidence that GCs do not satisfy the definition of a
galaxy---the iron abundances of their stars is explicable with only
stellar mass and Newtonian gravity.

Additional indirect diagnostics also demonstrate that GCs would be
classified as star clusters with our proposed definition.  The
presence of tidal streams around numerous MW GCs
\citep[e.g.][]{leon00} provides upper limits to their present-day
masses; this is additional evidence that their present-day dynamics are
consistent with their observed stellar masses and Newtonian gravity.
Another diagnostic is the existence of GCs in low-mass dwarf galaxies,
such as the Fornax dwarf spheroidal.  If its GCs were embedded in dark
matter halos, then their dynamical friction timescale for destruction
would be $<$1 Gyr, far shorter than their observed ages
\citep{conroy11a}.  One final diagnostic may be the outer density
profiles of GCs, as demonstrated by \citet{conroy11b} for the case of
NGC 2419 and MGC1.

Light element abundance spreads are common in GCs, and usually
attributed to enrichment by asymptotic giant branch stars or the winds
of rotating massive stars
\citep[e.g.,][]{renzini08a,ventura09a}. These ejecta are less
energetic than those of supernovae and can be retained by the gravity
of stars alone. The ubiquity of these abundance variations, often
identified through the anti-correlation of Na and O, has led to the
suggestion that such variations should \emph{define} the class of GCs
\citep{carretta10c}.  We do not advocate for this suggestion, since
little is known about the abundance patterns of low-mass GCs, which
may differ from those of more massive clusters, and star clusters with
masses $\la 10^{4} M_{\odot}$ in the Large Magellanic Cloud do not
appear to self-enrich \citep{milone09}. Furthermore, more massive
objects that might be confused with GCs, such as UCDs or dwarf nuclei,
lack detailed abundance observations.

Existing diagnostics do not preclude the hypothesis that some massive
($M_{V} <$ -10) GCs may reside in dark matter halos.  This
possibility must be considered when comparing observations against
cosmological models.  Extended star clusters ($M_{V} \sim -7$ to
$-8$, $r_{half} \sim$ 20 - 30 pc, \citealt{tanvir12a}), such as
those observed around M31 \citep{huxor05}, also present a challenge to
classification.  The M31 extended GCs would make particularly
interesting targets for spectroscopic [Fe/H] studies, because their
current stellar masses and escape velocities are too low to expect
self-enrichment in iron.

\subsection{Tidal Dwarfs}

The term ``tidal dwarf'' (TD) refers to a gravitationally bound,
galaxy-sized object (few kpc scale) formed as a result of the tidal
interaction of two galaxies \citep{bournaud10a}. These objects form
from a combination of star formation in gaseous tidal tails and of the
agglomeration of existing stars from the interacting parent galaxies
\citep{kaviraj12a}.  Candidates for such objects were originally
observed in the Antennae and in compact galaxy groups
\citep{mirabel92a,hunsberger96a}.  Although many candidate TDs have been
discovered since then, it remains difficult to determine whether TD
candidates are truly self-bound \citep{duc00a}.

Dynamical studies of TDs do not provide a definitive classification of
these objects.  Their kinematic properties are difficult to study, in
part because TDs are typically observed while still embedded in
ambient tidal material from which they formed/are forming.  Some
studies find their dynamical masses to be consistent with their
stellar and gas (both neutral and molecular) contents
\citep{duc00a,braine01a,bournaud04a,duc07a}, while others find
dynamical masses $2-3$ times higher than expected from observed stars
and gas \citep{bournaud07a}.  In all cases, the uncertainties are
substantial.  Even in a cold dark matter interpretation of galaxies,
TDs are not expected contain (much) dark matter
\citep[e.g.,][]{barnes92a}.  Unlike gas, the dark matter in TD
progenitor material cannot dissipate energy and has a velocity
dispersion exceeding the escape velocity of the forming TD
\citep{bournaud10a}, unless some dark matter is present in a cold,
rotating, galaxy disk \citep{purcell09a,read09a}.  Identifying a
sample of relatively older ($>$1 Gyr) TDs and conducting uniform
dynamical studies will help reveal whether: (i) TDs are simply
composed of gas and stars orbiting in a Newtownian potential, (ii)
galaxy disks do contain a dark matter component which can be accreted
by forming TDs, or (iii) TDs demonstrate a dynamical regime governed
by non-Newtonian gravity.  If (ii) or (iii) is verified, then TDs
would be classified by galaxies by our definition.

The possible contribution of ancient TDs formed at high redshift to
today's dwarf galaxy population, in particular around the Milky Way,
is controversial.  Observations of the universe at low or intermediate
redshift imply that that TDs could not contribute more than
$\sim$10\% of the dwarf galaxies in the local universe
\citep[e.g.,][]{wen12a,kaviraj12a}.  TDs forming in the local universe
also do not exhibit the relationship between stellar mass and
metallicity \citep{weilbacher03a} that is observed in the MW dwarfs
\citep{kirby11a}.  Moreover, kinematic studies of nearby TDs do not
imply the high dynamical $M/L$ observed for MW dwarf satellites.
Others propose that the MW's dwarf galaxies may be dominated by tidal
dwarfs formed at very high redshift when merger rates were far higher,
and that the high $M/L$ inferred for MW dwarfs are actually a
misinterpretation of the observed kinematics
\citep[e.g.][]{kuhn89,metz07,kroupa10a}.  It remains to be seen
whether models of ancient TDs evolving into z = 0 dwarfs could fall on
the same metallicity-luminosity relation followed by both MW dwarfs
and spheroidal galaxies over a wide range of masses.


\section{Discussion and Conclusions}

To compound the ambiguities inherent to classifying objects such as
extreme ultra-faint MW dwarfs and UCDs, observers have neither agreed
upon a definition of galaxy nor reached consensus on how to interpret
observations in hand.  To facilitate comparisons between dwarf galaxy
predictions and the increasingly complex sets of observations of
candidate dwarf galaxies, the field needs an agreed-upon definition
for galaxy.  We have accordingly proposed a physically motivated
definition that does not insist on a cold dark matter interpretation
of data: {\it A galaxy is a gravitationally bound collection of stars
  whose properties cannot be explained by a combination of baryons and
  Newton's laws of gravity.}

We have explored possible diagnostics of this galaxy definition
(primarily in the context of a cold dark matter dominated universe),
primarily kinematic studies and [Fe/H] spread.  Although kinematic
studies generally provide the most direct way to infer a galaxy
definition, it can be difficult to measure the dynamical mass of low
luminosity and/or low velocity dispersion ($<$ few \kms) systems.
Even once robustly measured, interpreting relatively modest dynamical
$M/L$ ($\lesssim 10$) may face several stumbling blocks: those that
could have generated overestimates of dynamical mass (e.g., binary
stars, contaminants in spectroscopic sample) and those could generate
underestimates of stellar mass (e.g., sparse sampling of the stellar
luminosity function, an overabundance of stellar remnants).  While
these effects do not appear to be a major problem for objects
currently classified as galaxies, including the Milky Way's dwarfs,
they should be carefully considered as discoveries at the extremes of
the cosmic zoo continue.  Systems such as UCDs and massive GCs also
may not bear a kinematic signature of dark matter or non-Newtonian
gravity, even if present, because their baryons are so densely packed.

$\sigma_{\rm [Fe/H]}$ provides an complimentary means to diagnose a
galaxy definition for systems less luminous than $M_{V} = -10$.  Using
public spectroscopic [Fe/H] measurements, we recalculated the average
systemic [Fe/H] and associated dispersions for 24 Milky Way GCs and 16
Milky Way dwarf galaxies.  All dwarf galaxies show spectroscopic
[Fe/H] spreads of $\sim$0.3 dex or more.  No GC less luminous than
$M_{V} = -10$ shows a notable ($\ga 0.1$ dex) [Fe/H] spread.  The
$\sigma_{\rm [Fe/H]}$ diagnostic has already been applied to the Segue
1 \citep{simon11a} and Willman 1 dwarf galaxies \citep{willman11a}.
One possible caveat with the $\sigma_{\rm [Fe/H]}$ diagnostic is the
possibility that the mergers of multiple star clusters could yield an
iron abundance spread.  This merging star cluster hypothesis, which
would produce a multimodal [Fe/H] distribution, should be carefully
considered when classifying objects by [Fe/H] dispersion alone.

The Fundamental Plane and its variants do not presently provide an
alternative means to diagnose a galaxy definition for low luminosity
systems.  However, these scaling relations do provide a useful
benchmark against which to compare ambiguous objects.  For example, an
outlier from known scaling relations may signal a problem with its
calculated velocity dispersion or estimated stellar mass (such as the
issues discussed in \S3.1.1.)  Well behaved scaling relations can also
help rule out pathological explanations for sets of objects,
especially when metallicity is included.  For example, the
metallicity--luminosity relation followed by the Milky Way's lowest
luminosity dwarfs helps rules out alternative hypotheses for their
existence as a population, such as tidal tails at apocenter and clumps
in streams.

 There are some classes of objects not discussed in this paper, but
 which would be worth new consideration in the context of our proposed
 galaxy definition. For example, dwarf ellipticals (dEs) do not
 typically show strong kinematic evidence for non-baryonic mass in
 their central regions \citep[e.g.][]{wolf10a,forbes11b}. However,
 recent kinematic studies of stars and globular clusters in their
 outer regions \citep{beasley09a,geha10a} have consistently suggested
 that $M/L$ increases with radius and that stars
 alone cannot account for the observations. Therefore, the current
 data favors the classification of dEs as galaxies by our
 definition. If future observations of dEs do not support this
 emerging consensus, then this classification should be revisited.

After examining massive globulars, UCDs and tidal dwarfs in detail, we
find that they can not yet conclusively be classified given existing
diagnostics of our galaxy definition.  Their ultimate classification
must be guided by future observational data.  If UCDs and tidal dwarfs
are inconsistent with a galaxy definition, this does not mean that
they should automatically be classified as star clusters.  Both of
these classes of objects are interesting and stand on their own as
worthy to investigate, given their unique properties and possibly
formation channels.  For tidal dwarfs, in particular, we advocate that
these objects are not lumped in with galaxies or clusters but that
they remain their own distinct class of objects.

We have suggested several measurements, some of which are possible
now, that could facilitate the classification of these, and other, extreme objects:\\

\noindent $\bullet$ Observational constraints on the [Fe/H] spread in
extended GCs and any UCD fainter than $M_{V} = -10$ (\S3.2.2,
\S4.3);\\

\noindent $\bullet$ Adaptive optics spectroscopy to measure the [Fe/H]
of individual stars in massive M31 GCs (\S3.2.3, \S4.3);\\

\noindent $\bullet$ Dynamical studies of the lowest luminosity
($M_{V} \gtrsim -9$), largest scale sized UCDs ($r_{half}
\gtrsim$ 30 pc) (\S4.2);\\

\noindent $\bullet$ Measuring dynamical masses of a larger sample of
UCDs to look for a positive correlation between half-light radius and
half-light $M/L$, at set UCD luminosity (\S4.2);\\

\noindent $\bullet$ Dynamical and chemical studies of extreme MW
 satellites Bo\"otes II and Segue 2 (\S4.1).\\

 Basic cold dark matter plus galaxy formation models predict a
 dichotomy between systems that form in the centers of dark matter
 halos and systems that form in the monolithic collapse of gas clouds
 that are not the primary baryonic components of dark matter
 halos. Our best present interpretation of the observations in this
 context reveals that systems forming at the center of a dark matter
 halo bear an observable imprint of this formation channel, such as
 the kinematic or chemical diagnostics discussed here.  This
 observable imprint would translate to a galaxy classification by our
 proposed definition.  The fact that systems classified as galaxies
 may be equivalent to the set of astrophysical systems that formed in
 dark matter halos can be used as strong guidance to theorists when
 selecting against which astrophysical systems to compare their
 predictions.  However, even in a dark matter context, we cannot take
 for granted that systems classified as galaxies by our definition are
 inclusive of all systems that formed in dark matter halos and
 exclusive of systems that formed otherwise.

 As our understanding of the universe grows, it may be possible for
 systems that formed inside of dark matter halos to fail the galaxy
 diagnostics discussed here.  For example, it would not be
 unreasonable to conceive that a very low-luminosity fossil galaxy
 could form all of its stars over a sufficiently short timescale that
 no opportunity for self-enrichment by supernovae could occur, leading
 to a minimal spread in [Fe/H].  If this is the case, then alternative
 diagnostics need to be identified for such ``first'' galaxies.  No
 objects meeting our definition of a galaxy via kinematics, but
 without a spread in [Fe/H], have yet been discovered, but it is
 plausible they exist.  Cosmological globular clusters, if they exist,
 may be such objects \citep{griffen10a}.  Conversely, it may be
 possible for stellar systems that formed inside of a dark matter halo
 to lose most or all of their dark matter.  In this scenario, a(n
 almost) stellar-only system may exist with the chemical imprint of
 formation within a dark matter halo.  Although simulations of dwarf
 galaxies in both cuspy and cored halos show that this is unlikely
 (see \S3.1), simulations of globular clusters within dark matter
 halos have shown that it may be possible to remove most of the dark
 matter in systems close to disruption \citep[e.g.][]{mashchenko05a}.
 
 Our proposed galaxy definition is itself independent of our
 observational knowledge and currently favored theories for structure
 formation; it can thus remain unchanged even as our understanding of
 the complex universe evolves.  However, the particular diagnostics of
 this definition as investigated in this polemic may indeed need to be
 revisited as our knowledge of extreme objects grows---both
 observationally and theoretically.  For example, the possible use of
 spreads in elements other than iron (such as calcium) to diagnose a
 galaxy classification is something that should continue to be
 scrutinized as our knowledge of such abundance patterns grow.


\acknowledgements

BW acknowledges support from NSF AST-0908193. BW also thanks NYU's
Center for Cosmology and Particle Physics and Drexel University's
Physics department for hosting her during the writing of of this
paper.  We thank Dr. Pierre-Alain Duc, the referee, for comments that
helped improve the quality and clarity of this paper.  We thank
Michele Bellazzini, Joerg Dabringhausen, Ross Fadely, Duncan Forbes,
Amanda Ford, Marla Geha, Amina Helmi, David Hogg, Evan Kirby, Pavel
Kroupa, George Lake, Erik Tollerud, and Enrico Vesperini for
stimulating conversations and emails leading up to and during the
preparation of this paper.  This research has made use of NASA's
Astrophysics Data System Bibliographic Services.


\clearpage

\begin{deluxetable}{lccccccccc}
  \tabletypesize{\scriptsize} \tablecaption{[Fe/H] properties of
    MW globular clusters and dwarfs} \tablewidth{0pt} 
\tablehead{
\colhead{Name} & 
\colhead{$[Fe/H]$} & 
\colhead{$\pm34\%$ CL} &
\colhead{$\sigma_{\rm [Fe/H]}$} & 
\colhead{+34\% CL} &
\colhead{-34\% CL} & 
\colhead{M$_{V}$} & 
\colhead{N$_{star}$} & 
\colhead{Ref} & 
\colhead{type} \\
\colhead{} & 
\colhead{dex} & 
\colhead{dex} &
\colhead{dex} & 
\colhead{dex} &
\colhead{dex} & 
\colhead{} & 
\colhead{} & 
\colhead{} & 
\colhead{}} 
\startdata
  $\omega$Cen & $-1.647$ & 0.009 & 0.271\tablenotemark{a} &0.007 &0.007 &$-10.3$ &	855	& J10 	&GC \\
  M54 	          & $-1.559$ & 0.021 & 0.186 & 0.016 &0.014 	&$-10.0$ &	76	&Car10 	&GC \\
  NGC 6441    & $-0.334$ & 0.018 &  0.079 &0.016 &0.013 	&$-9.6$ &	25	&G07 	&GC \\
  NGC 104      & $-0.743$ & 0.003 &  0.024 & 0.003 &0.002 	&$-9.4$ &	147	&Car09b 	&GC \\
  NGC 2419    & $-2.095$ & 0.019 & 0.032 & 0.013 &0.009 &$-9.4$ &	38	& Coh10 	&GC \\
  NGC 2808   & $-1.105$ & 0.006 &  0.062 & 0.005 &0.004 &$-9.4$ &	123	&Car06 	&GC \\
  NGC 6388   & $-0.404$ & 0.014 &  0.071 &0.012 &0.010 	&$-9.4$ &	36	&Car09b	&GC \\
  NGC 7078   & $-2.341$ & 0.007 &  0.055 &0.006 &0.005 	&$-9.2$ &	84	&Car09b 	&GC \\
  NGC 5904   & $-1.346$ & 0.002 &  0.014 &0.002 &0.002 	&$-8.8$ &	136	&Car09b 	&GC \\
  M22 	     & $-1.764$ & 0.016 & 0.099\tablenotemark{b}  &0.013 &0.011 	&$-8.5$ &	37	&M11 	&GC \\
  NGC 1851 & $-1.157$ & 0.005 & 0.046 & 0.004 & 0.003 & $-8.3$ & 124 & Car11 & GC \\
  NGC 1904  & $-1.545$ & 0.005 & 0.028 & 0.005 &0.004 & $-7.9$ &	58	&Car09b 	&GC \\
  NGC 6752 & $-1.564$ & 0.004 &  0.034 &0.003 &0.003 	& $-7.7$ &	137	&Car07b 	&GC \\
  NGC 6809  & $-1.970$ & 0.004 &  0.035 &0.003 &0.003 	&$-7.6$ &	156	&Car09b 	&GC \\
  NGC 3201  & $-1.495$ & 0.004 & 0.042 & 0.004 &0.003 	&$-7.5$ &	149	&Car09b 	&GC \\
  NGC 6254  & $-1.557$ & 0.005 &  0.048 &0.004 &0.003 	&$-7.5$ &	147	&Car09b 	&GC \\
  NGC 7099  & $-2.358$ & 0.006 & 0.037 &0.006 &0.005 	&$-7.5$ &	65	&Car09b 	&GC \\
  NGC 4590  & $-2.230$ & 0.007 &  0.057 &0.006 &0.005 	&$-7.4$ &	122	&Car09b 	&GC \\
  NGC 6218  & $-1.313$ & 0.004 & 0.027 &0.004 &0.003 	&$-7.3$ &	79	&Car07a	&GC \\
  NGC 6121  & $-1.200$ & 0.003 &  0.018 &0.003 &0.002 	&$-7.2$ &	103	&Car09b 	&GC \\
  NGC 6171  & $-1.066$ & 0.008 &  0.037 &0.007 &0.006 	&$-7.1$ &	33	&Car09b 	&GC \\
  NGC 288    & $-1.219$ & 0.004 & 0.034 & 0.004 &0.003 	&$-6.8$ &	110	&Car09b 	&GC \\
  NGC 6397 & $-1.994$ & 0.004 &  0.027 &0.003 &0.003 	&$-6.6$ &	144	&Car09b	&GC \\
  NGC 6838  & $-0.806$ & 0.006 &  0.027 &0.005 &0.005 	&$-5.6$ &	39	&Car09b	&GC \\
  For 	         & $-1.025$ & 0.012 & 0.292 &0.010 &0.010 	&$-13.3$ &	675	&K10 	&dwarf \\
  Leo I 	& $-1.450$ & 0.011 &  0.276 &0.009 &0.008 	&$-11.9$ &	827	&K10 	&dwarf \\
  Scl 	          & $-1.726$ & 0.024 &  0.452 &0.019 &0.017 	&$-11.2$ &	376	&K10 	&dwarf \\
  Leo II	& $-1.670$ & 0.024 &  0.347 &0.020 &0.018 	&$-10.0$ &	258	&K10 	&dwarf \\
  Sex 	& $-1.966$ & 0.039 &  0.339 &0.033 &0.030 	&$-9.6$ &	141	&K10	&dwarf \\
  Dra 	& $-1.946$ & 0.024 &  0.354 &0.020 &0.019 		&$-8.8$ &	298	&K10 	&dwarf \\
  CVn I 	& $-1.962$ & 0.038 &  0.441 &0.032 &0.029 	&$-8.6$ &	174	&K10 	&dwarf \\
  UMi 	& $-2.112$ & 0.027 &  0.319 &0.025 &0.023 	&$-9.2$ &	212	&K10 	&dwarf \\
  Herc 	& $-2.518$ & 0.140 &  0.583 &0.124 &0.095 	&$-6.2$ &	21	&K08 	&dwarf \\
  UMa I 	& $-2.334$ & 0.128 &  0.638 &0.106 &0.086 	&$-5.5$ &	31	&K08 	&dwarf \\
  Leo IV 	& $-2.363$ & 0.230 &  0.695 &0.210 &0.149 	&$-5.5$ &	12	&K08 	&dwarf \\
  Cvn II 	& $-2.444$ & 0.178 &  0.621 &0.164 &0.120 	&$-4.6$ &	15	&K08 	&dwarf \\
  UMa II 	& $-2.357$ & 0.204 &  0.563 &0.204 &0.136 	&$-4.0$ &	9	&K08 	&dwarf \\
  ComBer 	& $-2.640$ & 0.100 &  0.382 &0.088 &0.070 	&$-3.8$ &	23	&K08 	&dwarf \\
  Wil1 	& $-2.110$ & 0.367 &  0.557 &0.577 &0.231 	&$-2.7$ &	3	&W11 	&dwarf \\
  Seg 1      & $-2.735$ & 0.389\tablenotemark{c}  & 0.752 &0.417 &0.227 &$-1.5$ & 7 &N10, S11
  &dwarf\enddata 
\tablecomments{The reference column gives the source
    of individual [Fe/H] measurements used to estimate the dispersion
    in each object.  For Segue 1, only the one star (Seg 1-7) is taken
    from \citet{norris10a}.  Values of M$_{V}$ for the dwarfs are from
    \citet{sand11a} and references therein.  Values of M$_{V}$ for the
    GCs are from \citet[][2010 edition]{harrisGCcat}.  The posterior
    distribution of $[Fe/H]$ sufficiently symmetric that we only quote
    a single value for $\pm 34\%$ CL, taking the average of the + and
    $-$ values in the small number of cases with a few thousandth of a
    dex difference between the two..  Reference key: J10 =
    \citet{johnson10a}, Car11 = \citet{carretta11}, Car10 = \citet{carretta10b}, Coh10 =
    \citet{cohen10a}, Car09b = \citet{carretta09b}, Car06 =
    \citet{carretta06}, M11 = \citet{marino11a}, Car07a =
    \citet{carretta07a}, G07 = \citet{gratton07}, Car07b =
    \citet{carretta07b}, K10 = \citet{kirby10a}, K08 =
    \citet{kirby08a}, W11 = \citet{willman11a}, N10 =
    \citet{norris10a}, S11 = \citet{simon11a}}
\tablenotetext{a}{This value is a lower limit (see \S 3.2.1).} 
\tablenotetext{b}{This value is an upper limit (see \S 3.2.1).} 
\tablenotetext{c}{Unlike the other objects, the metallicity of Segue 1 has asymmetric uncertainties: $-2.735^{+0.373}_{-0.405}$} 

\end{deluxetable}


\clearpage

\bibliographystyle{apj} 

\begin{thebibliography}{155}
\expandafter\ifx\csname natexlab\endcsname\relax\def\natexlab#1{#1}\fi

\bibitem[{{Aihara}(2011)}]{dr8}
{Aihara}, H. e.~a. 2011, \apjs, 193, 29

\bibitem[{{Bailin} \& {Harris}(2009)}]{bailin09a}
{Bailin}, J., \& {Harris}, W.~E. 2009, \apj, 695, 1082

\bibitem[{{Barnes} \& {Hernquist}(1992)}]{barnes92a}
{Barnes}, J.~E., \& {Hernquist}, L. 1992, \nat, 360, 715

\bibitem[{{Baumgardt} {et~al.}(2008){Baumgardt}, {Kroupa}, \&
  {Parmentier}}]{baumgardt08b}
{Baumgardt}, H., {Kroupa}, P., \& {Parmentier}, G. 2008, \mnras, 384, 1231

\bibitem[{{Baumgardt} \& {Makino}(2003)}]{baumgardt03a}
{Baumgardt}, H., \& {Makino}, J. 2003, \mnras, 340, 227

\bibitem[{{Beasley} {et~al.}(2009){Beasley}, {Cenarro}, {Strader}, \&
  {Brodie}}]{beasley09a}
{Beasley}, M.~A., {Cenarro}, A.~J., {Strader}, J., \& {Brodie}, J.~P. 2009,
  \aj, 137, 5146

\bibitem[{{Bekki} \& {Freeman}(2003)}]{bekki03a}
{Bekki}, K., \& {Freeman}, K.~C. 2003, \mnras, 346, L11

\bibitem[{{Belokurov} {et~al.}(2009){Belokurov}, {Walker}, {Evans}, {Gilmore},
  {Irwin}, {Mateo}, {Mayer}, {Olszewski}, {Bechtold}, \&
  {Pickering}}]{belokurov09a}
{Belokurov}, V. {et~al.} 2009, \mnras, 397, 1748

\bibitem[{{Bournaud}(2010)}]{bournaud10a}
{Bournaud}, F. 2010, Advances in Astronomy, 2010

\bibitem[{{Bournaud} {et~al.}(2004){Bournaud}, {Duc}, {Amram}, {Combes}, \&
  {Gach}}]{bournaud04a}
{Bournaud}, F., {Duc}, P.-A., {Amram}, P., {Combes}, F., \& {Gach}, J.-L. 2004,
  \aap, 425, 813

\bibitem[{{Bournaud} {et~al.}(2007){Bournaud}, {Duc}, {Brinks}, {Boquien},
  {Amram}, {Lisenfeld}, {Koribalski}, {Walter}, \&
  {Charmandaris}}]{bournaud07a}
{Bournaud}, F. {et~al.} 2007, Science, 316, 1166

\bibitem[{{Boylan-Kolchin} {et~al.}(2011){Boylan-Kolchin}, {Bullock}, \&
  {Kaplinghat}}]{boylankolchin11a}
{Boylan-Kolchin}, M., {Bullock}, J.~S., \& {Kaplinghat}, M. 2011, \mnras, L267+

\bibitem[{{Boylan-Kolchin} {et~al.}(2012){Boylan-Kolchin}, {Bullock}, \&
  {Kaplinghat}}]{boylankolchin12a}
---. 2012, \mnras, 422, 1203

\bibitem[{{Bradford} {et~al.}(2011){Bradford}, {Geha}, {Mu{\~n}oz}, {Santana},
  {Simon}, {C{\^o}t{\'e}}, {Stetson}, {Kirby}, \& {Djorgovski}}]{bradford11a}
{Bradford}, J.~D. {et~al.} 2011, \apj, 743, 167

\bibitem[{{Braine} {et~al.}(2001){Braine}, {Duc}, {Lisenfeld}, {Charmandaris},
  {Vallejo}, {Leon}, \& {Brinks}}]{braine01a}
{Braine}, J., {Duc}, P.-A., {Lisenfeld}, U., {Charmandaris}, V., {Vallejo}, O.,
  {Leon}, S., \& {Brinks}, E. 2001, \aap, 378, 51

\bibitem[{{Brodie} {et~al.}(2011){Brodie}, {Romanowsky}, {Strader}, \&
  {Forbes}}]{brodie11a}
{Brodie}, J.~P., {Romanowsky}, A.~J., {Strader}, J., \& {Forbes}, D.~A. 2011,
  \aj, 142, 199

\bibitem[{{Brodie} \& {Strader}(2006)}]{brodie06a}
{Brodie}, J.~P., \& {Strader}, J. 2006, \araa, 44, 193

\bibitem[{{Busha} {et~al.}(2012){Busha}, {Lake}, \& {Reed}}]{busha12a}
{Busha}, R., {Lake}, G., \& {Reed}, D. 2012, in preparation

\bibitem[{{Carretta} {et~al.}(2009{\natexlab{a}}){Carretta}, {Bragaglia},
  {Gratton}, {D'Orazi}, \& {Lucatello}}]{carretta09a}
{Carretta}, E., {Bragaglia}, A., {Gratton}, R., {D'Orazi}, V., \& {Lucatello},
  S. 2009{\natexlab{a}}, \aap, 508, 695

\bibitem[{{Carretta} {et~al.}(2007{\natexlab{a}}){Carretta}, {Bragaglia},
  {Gratton}, {Catanzaro}, {Leone}, {Sabbi}, {Cassisi}, {Claudi}, {D'Antona},
  {Fran{\c c}ois}, {James}, \& {Piotto}}]{carretta07a}
{Carretta}, E. {et~al.} 2007{\natexlab{a}}, \aap, 464, 939

\bibitem[{{Carretta} {et~al.}(2006){Carretta}, {Bragaglia}, {Gratton}, {Leone},
  {Recio-Blanco}, \& {Lucatello}}]{carretta06}
{Carretta}, E., {Bragaglia}, A., {Gratton}, R.~G., {Leone}, F., {Recio-Blanco},
  A., \& {Lucatello}, S. 2006, \aap, 450, 523

\bibitem[{{Carretta} {et~al.}(2010{\natexlab{a}}){Carretta}, {Bragaglia},
  {Gratton}, {Lucatello}, {Bellazzini}, {Catanzaro}, {Leone}, {Momany},
  {Piotto}, \& {D'Orazi}}]{carretta10b}
{Carretta}, E. {et~al.} 2010{\natexlab{a}}, \aap, 520, A95

\bibitem[{{Carretta} {et~al.}(2009{\natexlab{b}}){Carretta}, {Bragaglia},
  {Gratton}, {Lucatello}, {Catanzaro}, {Leone}, {Bellazzini}, {Claudi},
  {D'Orazi}, {Momany}, {Ortolani}, {Pancino}, {Piotto}, {Recio-Blanco}, \&
  {Sabbi}}]{carretta09b}
---. 2009{\natexlab{b}}, \aap, 505, 117

\bibitem[{{Carretta} {et~al.}(2007{\natexlab{b}}){Carretta}, {Bragaglia},
  {Gratton}, {Lucatello}, \& {Momany}}]{carretta07b}
{Carretta}, E., {Bragaglia}, A., {Gratton}, R.~G., {Lucatello}, S., \&
  {Momany}, Y. 2007{\natexlab{b}}, \aap, 464, 927

\bibitem[{{Carretta} {et~al.}(2010{\natexlab{b}}){Carretta}, {Bragaglia},
  {Gratton}, {Recio-Blanco}, {Lucatello}, {D'Orazi}, \&
  {Cassisi}}]{carretta10c}
{Carretta}, E., {Bragaglia}, A., {Gratton}, R.~G., {Recio-Blanco}, A.,
  {Lucatello}, S., {D'Orazi}, V., \& {Cassisi}, S. 2010{\natexlab{b}}, \aap,
  516, A55

\bibitem[{{Carretta} {et~al.}(2011){Carretta}, {Lucatello}, {Gratton},
  {Bragaglia}, \& {D'Orazi}}]{carretta11}
{Carretta}, E., {Lucatello}, S., {Gratton}, R.~G., {Bragaglia}, A., \&
  {D'Orazi}, V. 2011, \aap, 533, A69

\bibitem[{{Chiboucas} {et~al.}(2011){Chiboucas}, {Tully}, {Marzke},
  {Phillipps}, {Price}, {Peng}, {Trentham}, {Carter}, \&
  {Hammer}}]{chiboucas11a}
{Chiboucas}, K. {et~al.} 2011, \apj, 737, 86

\bibitem[{{Chilingarian} {et~al.}(2011){Chilingarian}, {Mieske}, {Hilker}, \&
  {Infante}}]{chilingarian11a}
{Chilingarian}, I.~V., {Mieske}, S., {Hilker}, M., \& {Infante}, L. 2011,
  \mnras, 412, 1627

\bibitem[{{Cohen} {et~al.}(2010){Cohen}, {Kirby}, {Simon}, \&
  {Geha}}]{cohen10a}
{Cohen}, J.~G., {Kirby}, E.~N., {Simon}, J.~D., \& {Geha}, M. 2010, \apj, 725,
  288

\bibitem[{{Cohen} \& {Mel{\'e}ndez}(2005)}]{cohen05a}
{Cohen}, J.~G., \& {Mel{\'e}ndez}, J. 2005, \aj, 129, 303

\bibitem[{{Conroy} {et~al.}(2011){Conroy}, {Loeb}, \& {Spergel}}]{conroy11b}
{Conroy}, C., {Loeb}, A., \& {Spergel}, D.~N. 2011, \apj, 741, 72

\bibitem[{{Conroy} \& {Spergel}(2011)}]{conroy11a}
{Conroy}, C., \& {Spergel}, D.~N. 2011, \apj, 726, 36

\bibitem[{{Crommelin}(1918)}]{crommelin18a}
{Crommelin}, A.~C.~D. 1918, \jrasc, 12, 33

\bibitem[{{Da Rocha} {et~al.}(2011){Da Rocha}, {Mieske}, {Georgiev}, {Hilker},
  {Ziegler}, \& {Mendes de Oliveira}}]{darocha11a}
{Da Rocha}, C., {Mieske}, S., {Georgiev}, I.~Y., {Hilker}, M., {Ziegler},
  B.~L., \& {Mendes de Oliveira}, C. 2011, \aap, 525, A86

\bibitem[{{Dabringhausen} {et~al.}(2008){Dabringhausen}, {Hilker}, \&
  {Kroupa}}]{dabringhausen08a}
{Dabringhausen}, J., {Hilker}, M., \& {Kroupa}, P. 2008, \mnras, 386, 864

\bibitem[{{Dabringhausen} {et~al.}(2009){Dabringhausen}, {Kroupa}, \&
  {Baumgardt}}]{dabringhausen09a}
{Dabringhausen}, J., {Kroupa}, P., \& {Baumgardt}, H. 2009, \mnras, 394, 1529

\bibitem[{{Dabringhausen} {et~al.}(2012){Dabringhausen}, {Kroupa},
  {Pflamm-Altenburg}, \& {Mieske}}]{dabringhausen12a}
{Dabringhausen}, J., {Kroupa}, P., {Pflamm-Altenburg}, J., \& {Mieske}, S.
  2012, \apj, 747, 72

\bibitem[{{D'Antona} {et~al.}(2005){D'Antona}, {Bellazzini}, {Caloi}, {Pecci},
  {Galleti}, \& {Rood}}]{dantona05a}
{D'Antona}, F., {Bellazzini}, M., {Caloi}, V., {Pecci}, F.~F., {Galleti}, S.,
  \& {Rood}, R.~T. 2005, \apj, 631, 868

\bibitem[{{de Sitter}(1917)}]{desitter17}
{de Sitter}, W. 1917, \mnras, 78, 3

\bibitem[{{Dehnen} {et~al.}(2004){Dehnen}, {Odenkirchen}, {Grebel}, \&
  {Rix}}]{dehnen04}
{Dehnen}, W., {Odenkirchen}, M., {Grebel}, E.~K., \& {Rix}, H. 2004, \aj, 127,
  2753

\bibitem[{{Dekel} {et~al.}(2005){Dekel}, {Stoehr}, {Mamon}, {Cox}, {Novak}, \&
  {Primack}}]{dekel05a}
{Dekel}, A., {Stoehr}, F., {Mamon}, G.~A., {Cox}, T.~J., {Novak}, G.~S., \&
  {Primack}, J.~R. 2005, \nat, 437, 707

\bibitem[{{Dopita} \& {Smith}(1986)}]{dopita86}
{Dopita}, M.~A., \& {Smith}, G.~H. 1986, \apj, 304, 283

\bibitem[{{Drinkwater} {et~al.}(2003){Drinkwater}, {Gregg}, {Hilker}, {Bekki},
  {Couch}, {Ferguson}, {Jones}, \& {Phillipps}}]{drinkwater03a}
{Drinkwater}, M.~J., {Gregg}, M.~D., {Hilker}, M., {Bekki}, K., {Couch}, W.~J.,
  {Ferguson}, H.~C., {Jones}, J.~B., \& {Phillipps}, S. 2003, \nat, 423, 519

\bibitem[{{Duc}(2012)}]{duc12a}
{Duc}, P.-A. 2012, {Birth, Life and Survival of Tidal Dwarf Galaxies, in Dwarf
  Galaxies: Keys to Galaxy Formation and Evolution, Astrophysics and Space
  Science Proceedings, ISBN 978-3-642-22017-3.~Springer-Verlag Berlin
  Heidelberg, 2012, p.~305} (eds. {Papaderos}, P. and {Recchi}, S. and
  {Hensler}, G.)

\bibitem[{{Duc} \& {Bournaud}(2008)}]{duc08a}
{Duc}, P.-A., \& {Bournaud}, F. 2008, \apj, 673, 787

\bibitem[{{Duc} {et~al.}(2007){Duc}, {Braine}, {Lisenfeld}, {Brinks}, \&
  {Boquien}}]{duc07a}
{Duc}, P.-A., {Braine}, J., {Lisenfeld}, U., {Brinks}, E., \& {Boquien}, M.
  2007, \aap, 475, 187

\bibitem[{{Duc} {et~al.}(2000){Duc}, {Brinks}, {Springel}, {Pichardo},
  {Weilbacher}, \& {Mirabel}}]{duc00a}
{Duc}, P.-A., {Brinks}, E., {Springel}, V., {Pichardo}, B., {Weilbacher}, P.,
  \& {Mirabel}, I.~F. 2000, \aj, 120, 1238

\bibitem[{{Evstigneeva} {et~al.}(2007){Evstigneeva}, {Gregg}, {Drinkwater}, \&
  {Hilker}}]{evstigneeva07a}
{Evstigneeva}, E.~A., {Gregg}, M.~D., {Drinkwater}, M.~J., \& {Hilker}, M.
  2007, \aj, 133, 1722

\bibitem[{{Fadely} {et~al.}(2011){Fadely}, {Willman}, {Geha}, {Walsh},
  {Mu{\~n}oz}, {Jerjen}, {Vargas}, \& {Da Costa}}]{fadely11a}
{Fadely}, R., {Willman}, B., {Geha}, M., {Walsh}, S., {Mu{\~n}oz}, R.~R.,
  {Jerjen}, H., {Vargas}, L.~C., \& {Da Costa}, G.~S. 2011, \aj, 142, 88

\bibitem[{{Fellhauer} \& {Kroupa}(2005)}]{fellhauer05a}
{Fellhauer}, M., \& {Kroupa}, P. 2005, \mnras, 359, 223

\bibitem[{{Ferraro} {et~al.}(2009){Ferraro}, {Dalessandro}, {Mucciarelli},
  {Beccari}, {Rich}, {Origlia}, {Lanzoni}, {Rood}, {Valenti}, {Bellazzini},
  {Ransom}, \& {Cocozza}}]{ferraro09a}
{Ferraro}, F.~R. {et~al.} 2009, \nat, 462, 483

\bibitem[{{Firth} {et~al.}(2009){Firth}, {Evstigneeva}, \&
  {Drinkwater}}]{firth09a}
{Firth}, P., {Evstigneeva}, E.~A., \& {Drinkwater}, M.~J. 2009, \mnras, 394,
  1801

\bibitem[{{Forbes} \& {Kroupa}(2011)}]{forbes11a}
{Forbes}, D.~A., \& {Kroupa}, P. 2011, Publications of the Astronomical Society
  of Australia, 28, 77

\bibitem[{{Forbes} {et~al.}(2008){Forbes}, {Lasky}, {Graham}, \&
  {Spitler}}]{forbes08a}
{Forbes}, D.~A., {Lasky}, P., {Graham}, A.~W., \& {Spitler}, L. 2008, \mnras,
  389, 1924

\bibitem[{{Forbes} {et~al.}(2011){Forbes}, {Spitler}, {Graham}, {Foster},
  {Hau}, \& {Benson}}]{forbes11b}
{Forbes}, D.~A., {Spitler}, L.~R., {Graham}, A.~W., {Foster}, C., {Hau},
  G.~K.~T., \& {Benson}, A. 2011, \mnras, 413, 2665

\bibitem[{{Forbes} {et~al.}(2010){Forbes}, {Spitler}, {Harris}, {Bailin},
  {Strader}, {Brodie}, \& {Larsen}}]{forbes10b}
{Forbes}, D.~A., {Spitler}, L.~R., {Harris}, W.~E., {Bailin}, J., {Strader},
  J., {Brodie}, J.~P., \& {Larsen}, S.~S. 2010, \mnras, 403, 429

\bibitem[{{Frank} {et~al.}(2011){Frank}, {Hilker}, {Mieske}, {Baumgardt},
  {Grebel}, \& {Infante}}]{frank11a}
{Frank}, M.~J., {Hilker}, M., {Mieske}, S., {Baumgardt}, H., {Grebel}, E.~K.,
  \& {Infante}, L. 2011, \mnras, 414, L70

\bibitem[{{Fuentes-Carrera} {et~al.}(2008){Fuentes-Carrera}, {Jablonka},
  {Sarajedini}, {Bridges}, {Djorgovski}, \& {Meylan}}]{fuentescarrera08a}
{Fuentes-Carrera}, I., {Jablonka}, P., {Sarajedini}, A., {Bridges}, T.,
  {Djorgovski}, G., \& {Meylan}, G. 2008, \aap, 483, 769

\bibitem[{{Geha} {et~al.}(2010){Geha}, {van der Marel}, {Guhathakurta},
  {Gilbert}, {Kalirai}, \& {Kirby}}]{geha10a}
{Geha}, M., {van der Marel}, R.~P., {Guhathakurta}, P., {Gilbert}, K.~M.,
  {Kalirai}, J., \& {Kirby}, E.~N. 2010, \apj, 711, 361

\bibitem[{{Geha} {et~al.}(2009){Geha}, {Willman}, {Simon}, {Strigari}, {Kirby},
  {Law}, \& {Strader}}]{geha09a}
{Geha}, M., {Willman}, B., {Simon}, J.~D., {Strigari}, L.~E., {Kirby}, E.~N.,
  {Law}, D.~R., \& {Strader}, J. 2009, \apj, 692, 1464

\bibitem[{{Giersz}(2001)}]{giersz01a}
{Giersz}, M. 2001, \mnras, 324, 218

\bibitem[{{Gilmore} {et~al.}(2007){Gilmore}, {Wilkinson}, {Wyse}, {Kleyna},
  {Koch}, {Evans}, \& {Grebel}}]{gilmore07a}
{Gilmore}, G., {Wilkinson}, M.~I., {Wyse}, R.~F.~G., {Kleyna}, J.~T., {Koch},
  A., {Evans}, N.~W., \& {Grebel}, E.~K. 2007, \apj, 663, 948

\bibitem[{{Gnedin} {et~al.}(2002){Gnedin}, {Zhao}, {Pringle}, {Fall}, {Livio},
  \& {Meylan}}]{gnedin02a}
{Gnedin}, O.~Y., {Zhao}, H., {Pringle}, J.~E., {Fall}, S.~M., {Livio}, M., \&
  {Meylan}, G. 2002, \apjl, 568, L23

\bibitem[{{Governato} {et~al.}(2010){Governato}, {Brook}, {Mayer}, {Brooks},
  {Rhee}, {Wadsley}, {Jonsson}, {Willman}, {Stinson}, {Quinn}, \&
  {Madau}}]{governato10a}
{Governato}, F. {et~al.} 2010, \nat, 463, 203

\bibitem[{{Graham}(2011)}]{graham11a}
{Graham}, A.~W. 2011, ArXiv e-prints 1108.0997, condensed version of a review
  to appear in "Planets, Stars and Stellar Systems", Springer pub. 2012

\bibitem[{{Gratton} {et~al.}(2004){Gratton}, {Sneden}, \&
  {Carretta}}]{gratton04a}
{Gratton}, R., {Sneden}, C., \& {Carretta}, E. 2004, \araa, 42, 385

\bibitem[{{Gratton} {et~al.}(2007){Gratton}, {Lucatello}, {Bragaglia},
  {Carretta}, {Cassisi}, {Momany}, {Pancino}, {Valenti}, {Caloi}, {Claudi},
  {D'Antona}, {Desidera}, {Fran{\c c}ois}, {James}, {Moehler}, {Ortolani},
  {Pasquini}, {Piotto}, \& {Recio-Blanco}}]{gratton07}
{Gratton}, R.~G. {et~al.} 2007, \aap, 464, 953

\bibitem[{{Griffen} {et~al.}(2010){Griffen}, {Drinkwater}, {Thomas}, {Helly},
  \& {Pimbblet}}]{griffen10a}
{Griffen}, B.~F., {Drinkwater}, M.~J., {Thomas}, P.~A., {Helly}, J.~C., \&
  {Pimbblet}, K.~A. 2010, \mnras, 405, 375

\bibitem[{{Ha{\c s}egan} {et~al.}(2005){Ha{\c s}egan}, {Jord{\'a}n},
  {C{\^o}t{\'e}}, {Djorgovski}, {McLaughlin}, {Blakeslee}, {Mei}, {West},
  {Peng}, {Ferrarese}, {Milosavljevi{\'c}}, {Tonry}, \& {Merritt}}]{hasegan05a}
{Ha{\c s}egan}, M. {et~al.} 2005, \apj, 627, 203

\bibitem[{{Harris}(1996)}]{harrisGCcat}
{Harris}, W.~E. 1996, \aj, 112, 1487

\bibitem[{{Harris} {et~al.}(2006){Harris}, {Whitmore}, {Karakla}, {Oko{\'n}},
  {Baum}, {Hanes}, \& {Kavelaars}}]{harris06a}
{Harris}, W.~E., {Whitmore}, B.~C., {Karakla}, D., {Oko{\'n}}, W., {Baum},
  W.~A., {Hanes}, D.~A., \& {Kavelaars}, J.~J. 2006, \apj, 636, 90

\bibitem[{{Hernandez}(2012)}]{hernandez12a}
{Hernandez}, X. 2012, \mnras, 420, 1183

\bibitem[{{Hilker} {et~al.}(1999){Hilker}, {Infante}, {Vieira},
  {Kissler-Patig}, \& {Richtler}}]{hilker99a}
{Hilker}, M., {Infante}, L., {Vieira}, G., {Kissler-Patig}, M., \& {Richtler},
  T. 1999, \aaps, 134, 75

\bibitem[{{Hubble}(1926)}]{hubble26a}
{Hubble}, E.~P. 1926, \apj, 64, 321

\bibitem[{{Hunsberger} {et~al.}(1996){Hunsberger}, {Charlton}, \&
  {Zaritsky}}]{hunsberger96a}
{Hunsberger}, S.~D., {Charlton}, J.~C., \& {Zaritsky}, D. 1996, \apj, 462, 50

\bibitem[{{Huxor} {et~al.}(2005){Huxor}, {Tanvir}, {Irwin}, {Ibata}, {Collett},
  {Ferguson}, {Bridges}, \& {Lewis}}]{huxor05}
{Huxor}, A.~P., {Tanvir}, N.~R., {Irwin}, M.~J., {Ibata}, R., {Collett}, J.~L.,
  {Ferguson}, A.~M.~N., {Bridges}, T., \& {Lewis}, G.~F. 2005, \mnras, 360,
  1007

\bibitem[{{Ivezic} {et~al.}(2008){Ivezic}, {Tyson}, {Allsman}, {Andrew},
  {Angel}, \& {for the LSST Collaboration}}]{ivezic08a}
{Ivezic}, Z., {Tyson}, J.~A., {Allsman}, R., {Andrew}, J., {Angel}, R., \& {for
  the LSST Collaboration}. 2008, arXiv:0805.2366, also available at
  http://www.lsst.org/overview

\bibitem[{{Johnson} \& {Pilachowski}(2010)}]{johnson10a}
{Johnson}, C.~I., \& {Pilachowski}, C.~A. 2010, \apj, 722, 1373

\bibitem[{{Kaiser} {et~al.}(2002){Kaiser}, {Aussel}, {Burke}, {Boesgaard},
  {Chambers}, {Chun}, {Heasley}, {Hodapp}, {Hunt}, {Jedicke}, {Jewitt},
  {Kudritzki}, {Luppino}, {Maberry}, {Magnier}, {Monet}, {Onaka}, {Pickles},
  {Rhoads}, {Simon}, {Szalay}, {Szapudi}, {Tholen}, {Tonry}, {Waterson}, \&
  {Wick}}]{kaiser02a}
{Kaiser}, N. {et~al.} 2002, in Society of Photo-Optical Instrumentation
  Engineers (SPIE) Conference Series, Vol. 4836, Society of Photo-Optical
  Instrumentation Engineers (SPIE) Conference Series, ed. {J.~A.~Tyson \&
  S.~Wolff}, 154--164

\bibitem[{{Kaviraj} {et~al.}(2012){Kaviraj}, {Darg}, {Lintott}, {Schawinski},
  \& {Silk}}]{kaviraj12a}
{Kaviraj}, S., {Darg}, D., {Lintott}, C., {Schawinski}, K., \& {Silk}, J. 2012,
  \mnras, 419, 70

\bibitem[{{Keller} {et~al.}(2007){Keller}, {Schmidt}, {Bessell}, {Conroy},
  {Francis}, {Granlund}, {Kowald}, {Oates}, {Martin-Jones}, {Preston},
  {Tisserand}, {Vaccarella}, \& {Waterson}}]{keller07a}
{Keller}, S.~C. {et~al.} 2007, Publications of the Astronomical Society of
  Australia, 24, 1

\bibitem[{{Kirby} {et~al.}(2010){Kirby}, {Guhathakurta}, {Simon}, {Geha},
  {Rockosi}, {Sneden}, {Cohen}, {Sohn}, {Majewski}, \& {Siegel}}]{kirby10a}
{Kirby}, E.~N. {et~al.} 2010, \apjs, 191, 352

\bibitem[{{Kirby} {et~al.}(2011){Kirby}, {Lanfranchi}, {Simon}, {Cohen}, \&
  {Guhathakurta}}]{kirby11a}
{Kirby}, E.~N., {Lanfranchi}, G.~A., {Simon}, J.~D., {Cohen}, J.~G., \&
  {Guhathakurta}, P. 2011, \apj, 727, 78

\bibitem[{{Kirby} {et~al.}(2008){Kirby}, {Simon}, {Geha}, {Guhathakurta}, \&
  {Frebel}}]{kirby08a}
{Kirby}, E.~N., {Simon}, J.~D., {Geha}, M., {Guhathakurta}, P., \& {Frebel}, A.
  2008, \apjl, 685, L43

\bibitem[{{Klimentowski} {et~al.}(2007){Klimentowski}, {{\L}okas},
  {Kazantzidis}, {Prada}, {Mayer}, \& {Mamon}}]{klimentowski07}
{Klimentowski}, J., {{\L}okas}, E.~L., {Kazantzidis}, S., {Prada}, F., {Mayer},
  L., \& {Mamon}, G.~A. 2007, \mnras, 378, 353

\bibitem[{{Koposov} {et~al.}(2011){Koposov}, {Gilmore}, {Walker}, {Belokurov},
  {Wyn Evans}, {Fellhauer}, {Gieren}, {Geisler}, {Monaco}, {Norris}, {Okamoto},
  {Pe{\~n}arrubia}, {Wilkinson}, {Wyse}, \& {Zucker}}]{koposov11a}
{Koposov}, S.~E. {et~al.} 2011, \apj, 736, 146

\bibitem[{{Kroupa}(2008)}]{kroupa08a}
{Kroupa}, P. 2008, in IAU Symposium, Vol. 246, IAU Symposium, ed.
  {E.~Vesperini, M.~Giersz, \& A.~Sills}, 13--22

\bibitem[{{Kroupa} {et~al.}(2010){Kroupa}, {Famaey}, {de Boer},
  {Dabringhausen}, {Pawlowski}, {Boily}, {Jerjen}, {Forbes}, {Hensler}, \&
  {Metz}}]{kroupa10a}
{Kroupa}, P. {et~al.} 2010, \aap, 523, A32

\bibitem[{{Kuhn} \& {Miller}(1989)}]{kuhn89}
{Kuhn}, J.~R., \& {Miller}, R.~H. 1989, \apjl, 341, L41

\bibitem[{{Lee} {et~al.}(1999){Lee}, {Joo}, {Sohn}, {Rey}, {Lee}, \&
  {Walker}}]{lee99a}
{Lee}, Y., {Joo}, J., {Sohn}, Y., {Rey}, S., {Lee}, H., \& {Walker}, A.~R.
  1999, \nat, 402, 55

\bibitem[{{Leon} {et~al.}(2000){Leon}, {Meylan}, \& {Combes}}]{leon00}
{Leon}, S., {Meylan}, G., \& {Combes}, F. 2000, \aap, 359, 907

\bibitem[{{Madrid} {et~al.}(2010){Madrid}, {Graham}, {Harris}, {Goudfrooij},
  {Forbes}, {Carter}, {Blakeslee}, {Spitler}, \& {Ferguson}}]{madrid10a}
{Madrid}, J.~P. {et~al.} 2010, \apj, 722, 1707

\bibitem[{{Majewski} {et~al.}(2012){Majewski}, {Nidever}, {Smith}, {Damke},
  {Kunkel}, {Patterson}, {Bizyaev}, \& {Garc{\'{\i}}a P{\'e}rez}}]{maj12}
{Majewski}, S.~R., {Nidever}, D.~L., {Smith}, V.~V., {Damke}, G.~J., {Kunkel},
  W.~E., {Patterson}, R.~J., {Bizyaev}, D., \& {Garc{\'{\i}}a P{\'e}rez}, A.~E.
  2012, \apjl, 747, L37

\bibitem[{{Maraston} {et~al.}(2004){Maraston}, {Bastian}, {Saglia},
  {Kissler-Patig}, {Schweizer}, \& {Goudfrooij}}]{maraston04a}
{Maraston}, C., {Bastian}, N., {Saglia}, R.~P., {Kissler-Patig}, M.,
  {Schweizer}, F., \& {Goudfrooij}, P. 2004, \aap, 416, 467

\bibitem[{{Marino} {et~al.}(2011){Marino}, {Sneden}, {Kraft}, {Wallerstein},
  {Norris}, {da Costa}, {Milone}, {Ivans}, {Gonzalez}, {Fulbright}, {Hilker},
  {Piotto}, {Zoccali}, \& {Stetson}}]{marino11a}
{Marino}, A.~F. {et~al.} 2011, \aap, 532, A8

\bibitem[{{Martin} {et~al.}(2008){Martin}, {de Jong}, \& {Rix}}]{martin08b}
{Martin}, N.~F., {de Jong}, J.~T.~A., \& {Rix}, H.-W. 2008, \apj, 684, 1075

\bibitem[{{Martin} {et~al.}(2007){Martin}, {Ibata}, {Chapman}, {Irwin}, \&
  {Lewis}}]{martin07a}
{Martin}, N.~F., {Ibata}, R.~A., {Chapman}, S.~C., {Irwin}, M., \& {Lewis},
  G.~F. 2007, \mnras, 380, 281

\bibitem[{{Martinez} {et~al.}(2011){Martinez}, {Minor}, {Bullock},
  {Kaplinghat}, {Simon}, \& {Geha}}]{martinez11a}
{Martinez}, G.~D., {Minor}, Q.~E., {Bullock}, J., {Kaplinghat}, M., {Simon},
  J.~D., \& {Geha}, M. 2011, \apj, 738, 55

\bibitem[{{Mashchenko} \& {Sills}(2005)}]{mashchenko05a}
{Mashchenko}, S., \& {Sills}, A. 2005, \apj, 619, 258

\bibitem[{{McConnachie} \& {C{\^o}t{\'e}}(2010)}]{mcconnachie10a}
{McConnachie}, A.~W., \& {C{\^o}t{\'e}}, P. 2010, \apjl, 722, L209

\bibitem[{{Messier}(1781)}]{messier1781}
{Messier}, C. 1781, Connoissance des Temps for 1784

\bibitem[{{Metz} \& {Kroupa}(2007)}]{metz07}
{Metz}, M., \& {Kroupa}, P. 2007, \mnras, 376, 387

\bibitem[{{Meylan} {et~al.}(2001){Meylan}, {Sarajedini}, {Jablonka},
  {Djorgovski}, {Bridges}, \& {Rich}}]{meylan01a}
{Meylan}, G., {Sarajedini}, A., {Jablonka}, P., {Djorgovski}, S.~G., {Bridges},
  T., \& {Rich}, R.~M. 2001, \aj, 122, 830

\bibitem[{{Mieske} {et~al.}(2008){Mieske}, {Hilker}, {Jord{\'a}n}, {Infante},
  {Kissler-Patig}, {Rejkuba}, {Richtler}, {C{\^o}t{\'e}}, {Baumgardt}, {West},
  {Ferrarese}, \& {Peng}}]{mieske08a}
{Mieske}, S. {et~al.} 2008, \aap, 487, 921

\bibitem[{{Mieske} {et~al.}(2006){Mieske}, {Jord{\'a}n}, {C{\^o}t{\'e}},
  {Kissler-Patig}, {Peng}, {Ferrarese}, {Blakeslee}, {Mei}, {Merritt}, {Tonry},
  \& {West}}]{mieske06a}
---. 2006, \apj, 653, 193

\bibitem[{{Mieske} {et~al.}(2010){Mieske}, {Jord{\'a}n}, {C{\^o}t{\'e}},
  {Peng}, {Ferrarese}, {Blakeslee}, {Mei}, {Baumgardt}, {Tonry}, {Infante}, \&
  {West}}]{mieske10a}
---. 2010, \apj, 710, 1672

\bibitem[{{Milgrom}(1983)}]{milgrom83}
{Milgrom}, M. 1983, \apj, 270, 371

\bibitem[{{Milone} {et~al.}(2009){Milone}, {Bedin}, {Piotto}, \&
  {Anderson}}]{milone09}
{Milone}, A.~P., {Bedin}, L.~R., {Piotto}, G., \& {Anderson}, J. 2009, \aap,
  497, 755

\bibitem[{{Minchin} {et~al.}(2005){Minchin}, {Davies}, {Disney}, {Boyce},
  {Garcia}, {Jordan}, {Kilborn}, {Lang}, {Roberts}, {Sabatini}, \& {van
  Driel}}]{minchin05}
{Minchin}, R. {et~al.} 2005, \apjl, 622, L21

\bibitem[{{Mirabel} {et~al.}(1992){Mirabel}, {Dottori}, \& {Lutz}}]{mirabel92a}
{Mirabel}, I.~F., {Dottori}, H., \& {Lutz}, D. 1992, \aap, 256, L19

\bibitem[{{Misgeld} \& {Hilker}(2011)}]{misgeld11a}
{Misgeld}, I., \& {Hilker}, M. 2011, \mnras, 414, 3699

\bibitem[{{Misgeld} {et~al.}(2011){Misgeld}, {Mieske}, {Hilker}, {Richtler},
  {Georgiev}, \& {Schuberth}}]{misgeld11b}
{Misgeld}, I., {Mieske}, S., {Hilker}, M., {Richtler}, T., {Georgiev}, I.~Y.,
  \& {Schuberth}, Y. 2011, \aap, 531, A4

\bibitem[{{Moore} {et~al.}(2006){Moore}, {Diemand}, {Madau}, {Zemp}, \&
  {Stadel}}]{moore06a}
{Moore}, B., {Diemand}, J., {Madau}, P., {Zemp}, M., \& {Stadel}, J. 2006,
  \mnras, 368, 563

\bibitem[{{Mu{\~n}oz} {et~al.}(2010){Mu{\~n}oz}, {Geha}, \&
  {Willman}}]{munoz10a}
{Mu{\~n}oz}, R.~R., {Geha}, M., \& {Willman}, B. 2010, \aj, 140, 138

\bibitem[{{Norris} {et~al.}(2010){Norris}, {Wyse}, {Gilmore}, {Yong}, {Frebel},
  {Wilkinson}, {Belokurov}, \& {Zucker}}]{norris10a}
{Norris}, J.~E., {Wyse}, R.~F.~G., {Gilmore}, G., {Yong}, D., {Frebel}, A.,
  {Wilkinson}, M.~I., {Belokurov}, V., \& {Zucker}, D.~B. 2010, \apj, 723, 1632

\bibitem[{{Norris} \& {Kannappan}(2011)}]{norris11b}
{Norris}, M.~A., \& {Kannappan}, S.~J. 2011, \mnras, 414, 739

\bibitem[{{Odenkirchen} {et~al.}(2002){Odenkirchen}, {Grebel}, {Dehnen}, {Rix},
  \& {Cudworth}}]{odenkirchen02a}
{Odenkirchen}, M., {Grebel}, E.~K., {Dehnen}, W., {Rix}, H.-W., \& {Cudworth},
  K.~M. 2002, \aj, 124, 1497

\bibitem[{{Origlia} {et~al.}(2011){Origlia}, {Rich}, {Ferraro}, {Lanzoni},
  {Bellazzini}, {Dalessandro}, {Mucciarelli}, {Valenti}, \&
  {Beccari}}]{origlia11a}
{Origlia}, L. {et~al.} 2011, \apjl, 726, L20

\bibitem[{{Pe{\~n}arrubia} {et~al.}(2010){Pe{\~n}arrubia}, {Benson}, {Walker},
  {Gilmore}, {McConnachie}, \& {Mayer}}]{penarrubia10a}
{Pe{\~n}arrubia}, J., {Benson}, A.~J., {Walker}, M.~G., {Gilmore}, G.,
  {McConnachie}, A.~W., \& {Mayer}, L. 2010, \mnras, 406, 1290

\bibitem[{{Pe{\~n}arrubia} {et~al.}(2008){Pe{\~n}arrubia}, {Navarro}, \&
  {McConnachie}}]{penarrubia08a}
{Pe{\~n}arrubia}, J., {Navarro}, J.~F., \& {McConnachie}, A.~W. 2008, \apj,
  673, 226

\bibitem[{{Pontzen} \& {Governato}(2012)}]{pontzen12a}
{Pontzen}, A., \& {Governato}, F. 2012, \mnras, 421, 3464

\bibitem[{{Purcell} {et~al.}(2009){Purcell}, {Bullock}, \&
  {Kaplinghat}}]{purcell09a}
{Purcell}, C.~W., {Bullock}, J.~S., \& {Kaplinghat}, M. 2009, \apj, 703, 2275

\bibitem[{{Read} {et~al.}(2009){Read}, {Mayer}, {Brooks}, {Governato}, \&
  {Lake}}]{read09a}
{Read}, J.~I., {Mayer}, L., {Brooks}, A.~M., {Governato}, F., \& {Lake}, G.
  2009, \mnras, 397, 44

\bibitem[{{Renzini}(2008)}]{renzini08a}
{Renzini}, A. 2008, \mnras, 391, 354

\bibitem[{{Romanowsky} {et~al.}(2003){Romanowsky}, {Douglas}, {Arnaboldi},
  {Kuijken}, {Merrifield}, {Napolitano}, {Capaccioli}, \&
  {Freeman}}]{romanowsky03a}
{Romanowsky}, A.~J., {Douglas}, N.~G., {Arnaboldi}, M., {Kuijken}, K.,
  {Merrifield}, M.~R., {Napolitano}, N.~R., {Capaccioli}, M., \& {Freeman},
  K.~C. 2003, Science, 301, 1696

\bibitem[{{Sand} {et~al.}(2011){Sand}, {Strader}, {Willman}, {Zaritsky},
  {McLeod}, {Caldwell}, {Seth}, \& {Olszewski}}]{sand11a}
{Sand}, D.~J., {Strader}, J., {Willman}, B., {Zaritsky}, D., {McLeod}, B.,
  {Caldwell}, N., {Seth}, A., \& {Olszewski}, E. 2011, ArXiv e-prints 1111.6608

\bibitem[{{Sarajedini} \& {Layden}(1995)}]{sarajedini95a}
{Sarajedini}, A., \& {Layden}, A.~C. 1995, \aj, 109, 1086

\bibitem[{{Saviane} {et~al.}(2012){Saviane}, {Da Costa}, {Held}, {Sommariva},
  {Gullieuszik}, {Barbuy}, \& {Ortolani}}]{saviane12a}
{Saviane}, I., {Da Costa}, G.~S., {Held}, E.~V., {Sommariva}, V.,
  {Gullieuszik}, M., {Barbuy}, B., \& {Ortolani}, S. 2012, ArXiv e-prints

\bibitem[{{Shapley}(1919)}]{shapley19a}
{Shapley}, H. 1919, \pasp, 31, 261

\bibitem[{{Simon} \& {Geha}(2007)}]{simon07a}
{Simon}, J.~D., \& {Geha}, M. 2007, \apj, 670, 313

\bibitem[{{Simon} {et~al.}(2011){Simon}, {Geha}, {Minor}, {Martinez}, {Kirby},
  {Bullock}, {Kaplinghat}, {Strigari}, {Willman}, {Choi}, {Tollerud}, \&
  {Wolf}}]{simon11a}
{Simon}, J.~D. {et~al.} 2011, \apj, 733, 46

\bibitem[{{Skillman} {et~al.}(1989){Skillman}, {Kennicutt}, \&
  {Hodge}}]{skillman89a}
{Skillman}, E.~D., {Kennicutt}, R.~C., \& {Hodge}, P.~W. 1989, \apj, 347, 875

\bibitem[{{Slipher}(1917)}]{slipher17a}
{Slipher}, V.~M. 1917, Proceedings of the American Philosophical Society, 56,
  403

\bibitem[{{Sotiriou} \& {Faraoni}(2010)}]{sotiriou10}
{Sotiriou}, T.~P., \& {Faraoni}, V. 2010, Reviews of Modern Physics, 82, 451

\bibitem[{{Spitler} {et~al.}(2006){Spitler}, {Larsen}, {Strader}, {Brodie},
  {Forbes}, \& {Beasley}}]{spitler06a}
{Spitler}, L.~R., {Larsen}, S.~S., {Strader}, J., {Brodie}, J.~P., {Forbes},
  D.~A., \& {Beasley}, M.~A. 2006, \aj, 132, 1593

\bibitem[{{Strader} {et~al.}(2006){Strader}, {Brodie}, {Spitler}, \&
  {Beasley}}]{strader06a}
{Strader}, J., {Brodie}, J.~P., {Spitler}, L., \& {Beasley}, M.~A. 2006, \aj,
  132, 2333

\bibitem[{{Strader} {et~al.}(2011){Strader}, {Caldwell}, \&
  {Seth}}]{strader11a}
{Strader}, J., {Caldwell}, N., \& {Seth}, A.~C. 2011, \aj, 142, 8

\bibitem[{{Strader} \& {Smith}(2008)}]{strader08a}
{Strader}, J., \& {Smith}, G.~H. 2008, \aj, 136, 1828

\bibitem[{{Tanvir} {et~al.}(2012){Tanvir}, {Mackey}, {Ferguson}, {Huxor},
  {Read}, {Lewis}, {Irwin}, {Chapman}, {Ibata}, {Wilkinson}, {McConnachie},
  {Martin}, {Davies}, \& {Bridges}}]{tanvir12a}
{Tanvir}, N.~R. {et~al.} 2012, \mnras, 422, 162

\bibitem[{{The Dark Energy Survey Collaboration}(2005)}]{DES}
{The Dark Energy Survey Collaboration}. 2005, White Paper submitted to the Dark
  Energy Task Force, arXiv:astro-ph/0510346

\bibitem[{{Tollerud} {et~al.}(2011){Tollerud}, {Bullock}, {Graves}, \&
  {Wolf}}]{tollerud11a}
{Tollerud}, E.~J., {Bullock}, J.~S., {Graves}, G.~J., \& {Wolf}, J. 2011, \apj,
  726, 108

\bibitem[{{Tremonti} {et~al.}(2004){Tremonti}, {Heckman}, {Kauffmann},
  {Brinchmann}, {Charlot}, {White}, {Seibert}, {Peng}, {Schlegel}, {Uomoto},
  {Fukugita}, \& {Brinkmann}}]{tremonti04}
{Tremonti}, C.~A. {et~al.} 2004, \apj, 613, 898

\bibitem[{{van den Bergh}(2008)}]{vandenbergh08a}
{van den Bergh}, S. 2008, \mnras, 385, L20

\bibitem[{{Ventura} \& {D'Antona}(2009)}]{ventura09a}
{Ventura}, P., \& {D'Antona}, F. 2009, \aap, 499, 835

\bibitem[{{Vesperini} \& {Heggie}(1997)}]{vesperini97}
{Vesperini}, E., \& {Heggie}, D.~C. 1997, \mnras, 289, 898

\bibitem[{{Walker} {et~al.}(2009{\natexlab{a}}){Walker}, {Mateo}, {Olszewski},
  {Pe{\~n}arrubia}, {Wyn Evans}, \& {Gilmore}}]{walker09c}
{Walker}, M.~G., {Mateo}, M., {Olszewski}, E.~W., {Pe{\~n}arrubia}, J., {Wyn
  Evans}, N., \& {Gilmore}, G. 2009{\natexlab{a}}, \apj, 704, 1274

\bibitem[{{Walker} {et~al.}(2009{\natexlab{b}}){Walker}, {Mateo}, {Olszewski},
  {Sen}, \& {Woodroofe}}]{walker09b}
{Walker}, M.~G., {Mateo}, M., {Olszewski}, E.~W., {Sen}, B., \& {Woodroofe}, M.
  2009{\natexlab{b}}, \aj, 137, 3109

\bibitem[{{Walsh} {et~al.}(2008){Walsh}, {Willman}, {Sand}, {Harris}, {Seth},
  {Zaritsky}, \& {Jerjen}}]{walsh08a}
{Walsh}, S.~M., {Willman}, B., {Sand}, D., {Harris}, J., {Seth}, A.,
  {Zaritsky}, D., \& {Jerjen}, H. 2008, \apj, 688, 245

\bibitem[{{Weilbacher} {et~al.}(2003){Weilbacher}, {Duc}, \&
  {Fritze-v.~Alvensleben}}]{weilbacher03a}
{Weilbacher}, P.~M., {Duc}, P.-A., \& {Fritze-v.~Alvensleben}, U. 2003, \aap,
  397, 545

\bibitem[{{Wen} {et~al.}(2012){Wen}, {Zheng}, {Zhao}, \& {Gao}}]{wen12a}
{Wen}, Z.-Z., {Zheng}, X.-Z., {Zhao}, Y.-H., \& {Gao}, Y. 2012, \apss, 337, 729

\bibitem[{{Willman} {et~al.}(2011){Willman}, {Geha}, {Strader}, {Strigari},
  {Simon}, {Kirby}, {Ho}, \& {Warres}}]{willman11a}
{Willman}, B., {Geha}, M., {Strader}, J., {Strigari}, L.~E., {Simon}, J.~D.,
  {Kirby}, E., {Ho}, N., \& {Warres}, A. 2011, \aj, 142, 128

\bibitem[{{Wolf} {et~al.}(2010){Wolf}, {Martinez}, {Bullock}, {Kaplinghat},
  {Geha}, {Mu{\~n}oz}, {Simon}, \& {Avedo}}]{wolf10a}
{Wolf}, J., {Martinez}, G.~D., {Bullock}, J.~S., {Kaplinghat}, M., {Geha}, M.,
  {Mu{\~n}oz}, R.~R., {Simon}, J.~D., \& {Avedo}, F.~F. 2010, \mnras, 406, 1220

\bibitem[{{Woo} {et~al.}(2008){Woo}, {Courteau}, \& {Dekel}}]{woo08a}
{Woo}, J., {Courteau}, S., \& {Dekel}, A. 2008, \mnras, 390, 1453

\bibitem[{{Zaritsky} {et~al.}(2011){Zaritsky}, {Zabludoff}, \&
  {Gonzalez}}]{zaritsky11a}
{Zaritsky}, D., {Zabludoff}, A.~I., \& {Gonzalez}, A.~H. 2011, \apj, 727, 116

\bibitem[{{Zolotov} {et~al.}(2011){Zolotov}, {Hogg}, \& {Willman}}]{zolotov11a}
{Zolotov}, A., {Hogg}, D.~W., \& {Willman}, B. 2011, \apjl, 727, L14

\end{thebibliography}

\end{document}